\documentclass[usenatbib]{mn2e}
\usepackage{natbibmnfix,graphicx,times, amsmath}

\def\myputfigure#1#2#3#4#5%
{\vskip#5pt\makebox[0pt]{\hskip#2in
\includegraphics[width=#3\textwidth]{#1}}\vskip#4pt\hfill}

\newcommand\lsim{\mathrel{\rlap{\lower4pt\hbox{\hskip1pt$\sim$}}
        \raise1pt\hbox{$<$}}}
\newcommand\gsim{\mathrel{\rlap{\lower4pt\hbox{\hskip1pt$\sim$}}
        \raise1pt\hbox{$>$}}}

\newcommand{\nf}{x_{\rm HI}}

\newcommand{\Msun}{M_\odot}
\newcommand{\Tvir}{T_{\rm vir}}

\newcommand{\zinit}{z_{\rm init}}

\newcommand{\fcd}{f_{\rm cd}}

\newcommand{\flash}{Flash}
\newcommand{\nothing}{NoUVB}
\newcommand{\heatinglow}{Heat0.08}
\newcommand{\heatinghigh}{Heat0.8}
\newcommand{\LWfournothing}{LW10$^{-4}$\_NoUVB}
\newcommand{\LWfourheatinghigh}{LW10$^{-4}$\_Heat0.8}
\newcommand{\LWthreenothing}{LW10$^{-3}$\_NoUVB}
\newcommand{\LWthreeheatinghigh}{LW10$^{-3}$\_Heat0.8}
\newcommand{\LWtwonothing}{LW10$^{-2}$\_NoUVB}
\newcommand{\LWtwoheatinghigh}{LW10$^{-2}$\_Heat0.8}
\newcommand{\LWonenothing}{LW0.1\_NoUVB}
\newcommand{\LWoneheatinghigh}{LW0.1\_Heat0.8}
\newcommand{\Juvb}{J_{21}^{UV}}
\newcommand{\zuvbon}{z_{\rm UVB, on}}
\newcommand{\zuvboff}{z_{\rm UVB, off}}
\newcommand{\delN}{\delta_{N, {\rm cd}}(z)}
\newcommand{\delM}{\delta_{M, {\rm cd}}(z)}
\newcommand{\Nrun}{N_{\rm cd}^{{\rm run}i}}
\newcommand{\Mrun}{M_{\rm cd}^{{\rm run}i}}

\newcommand{\hMpc}{h^{-1} ~ {\rm Mpc}}

\newcommand{\tHtwo}{t_{\rm H_2}}

\newcommand{\fdelay}{f_{\rm delay}}
\newcommand{\JLW}{J_{\rm LW}}
\newcommand{\Jlwb}{J_{21}^{\rm LW}}

\begin{document}


\title[High-z Relic HII Regions and Radiative Feedback]{Relic HII Regions and Radiative Feedback at High-Redshifts}

\author[Mesinger et al.]{Andrei Mesinger$^1$\thanks{Hubble Fellow; e-mail: mesinger@astro.princeton.edu}, Greg L. Bryan$^2$, \& Zolt\'{a}n Haiman$^2$ \\
$^1$Department of Astrophysical Sciences, Princeton University, Princeton, NJ 08544, USA\\
$^2$Department of Astronomy, Columbia University, 550 West 120th  
Street, New York, NY 10027, USA}

\voffset-.6in

\maketitle

\begin{abstract}
UV radiation from early astrophysical sources could have a large impact on subsequent star formation in nearby protogalaxies, and in general on the progress of cosmological reionization.  Theoretical arguments based on the absence of metals in the early Universe suggest that the first stars were likely massive, bright, yet short-lived, with lifetimes of a few million years.  Here we study the radiative feedback arising from such stars using hydrodynamical simulations with transient UV backgrounds (UVBs) and persistent Lyman-Werner backgrounds (LWBs) of varying intensity.  We extend our prior work in Mesinger et al. (2006), by studying a more typical region whose proto-galaxies form at lower redshifts, $z\sim$ 13--20, in the epoch likely preceding the bulk of reionization.
We confirm our previous results that feedback in the relic HII regions resulting from such transient radiation, is itself transient.  Feedback effects dwindle away after $\sim30\%$ of the Hubble time, and the same critical specific intensity of $J_{\rm UV} \sim 0.1 \times 10^{-21}{\rm ergs~s^{-1}~cm^{-2}~Hz^{-1}~sr^{-1}}$ separates positive and negative feedback regimes. This suggests that overall feedback is fairly  insensitive to the large-scale environment, overdensity, and redshift-dependent halo parameters, and can accurately be modeled in this regime with just the intensity of the impinging UVB.
Additionally, we discover a second episode of eventual positive feedback in halos which have not yet collapsed when their progenitor regions were exposed to the transient UVB.  When exposed to the transient UVB, this gas suffers relatively little density depletion but a significant enhancement of the molecular hydrogen abundance, thus resulting in net positive feedback.  This eventual positive feedback appears in all runs, regardless of the strength of the UVB.
However, this feedback regime is very sensitive to the presence of Lyman-Werner radiation, and notable effects disappear under fairly modest background intensities of $\JLW \gsim 10^{-3} \times 10^{-21}{\rm ergs~s^{-1}~cm^{-2}~Hz^{-1}~sr^{-1}}$, assuming the region is optically thin for LW photons. Nevertheless, when exposed to the same LWB, halos inside relic HII regions always have a higher H$_2$ abundance and shorter cooling times than halos outside relic HII regions, allowing gas to cool faster once it finally begins to collapse onto the halo.
We conclude that UV radiative feedback in relic HII regions, although a complicated process, seems unlikely to have a major impact on the progress of cosmological reionization, provided that present estimates of the lifetime and luminosity of a Pop III star are accurate.  More likely is that the build-up of the LWB ultimately governs the feedback strength until a persistent UV background can be established.
\end{abstract}


\section{Introduction}
\label{sec:intro}

Semi-analytic models and numerical simulations predict that the first astrophysical objects formed at redshifts $z\gsim$ 20.   They form out of metal-free gas with inefficient fragmentation, thus numerical simulations predict that these first objects could be very massive ($M_\ast\gsim 100 \Msun$), so-called 'Population III' (Pop III) stars \citep{ABN02, BCL02, YOH08}.  These stars would be short-lived, with lifetimes of $\sim 3$ Myr, but could produce ten times more ionizing photons per baryon than regular Population II stars \citep{Schaerer02, Schaerer03}.  Also, such star formation could have been somewhat synchronized, given that these early objects formed in highly biased and clustered environments.  Hence, Pop III stars could have a significant impact on subsequent generations of objects.

Indeed Pop III stars can strongly affect the progress of cosmological reionization \citep{HH03, Cen03a, Cen03b, WL03_postWMAP}, although their contribution is virtually unknown from first principles and is generally applied to simulations with a ``tuning knob" approach.  Thus ionizing photons from the first stars and lack thereof have been used to explain both early (e.g. \citealt{Cen03b}) and fairly late (e.g. \citealt{HB06}) reionization, depending on which scenario was favored when the works were published.  Therefore, it is quite important to gain physical insight into how the radiation of the first stars impacted their surroundings.  

Radiative feedback can be either positive or negative, in that it can
enhance or suppress subsequent star--formation.  Positive feedback can
result when the enhanced free-electron fraction from ionizing photons (e.g. \citealt{OH02}) or hydrodynamical shocks \citep{SK87}
catalyzes the formation of molecular hydrogen (H$_2$). If the ionizing background ``turns-off" (resulting in so-called ``relic" HII regions), H$_2$ cooling can provide the dominant cooling channel at high-densities and low temperatures.
Conversely, negative feedback can result from heating by ionizing radiation which can photo-evaporate gas in low-mass halos \citep{Efstathiou92, BL99, Gnedin00filter, SIR04, Dijkstra04, MD08}.  Also, an active background of Lyman-Werner (LW) radiation (with photon energies in the 11.18--13.6eV range) can dissociate H$_2$, thus decreasing the gas's cooling capabilities (e.g., \citealt{HRL97, HAR00, CFA00, MBA01, WA07, Oshea08}).

Indeed, simulations find radiative feedback to be nuanced at very high redshifts.  Positive feedback dominates in flash ionized gas \citep{OShea05}, or gas exposed to a weak transient ultraviolet background (UVB) (\citealt{MBH06}, hereafter MBH06) such as might be present close to the edges of HII regions \citep{RGS02b, KM05}.  On the other hand, negative feedback in relic HII regions can occur in regions closer to the Pop III star (MBH06; \citealt{SU06, AS07, Yoshida07}).  Furthermore, radiative transfer effects can also impact the strength and sign of feedback (e.g. \citealt{ISR05, SU06, AWB07, Whalen08, WA08}).  Because of variations in halo mass, collapse redshift, central baryon density, stellar spectrum, and distance to the ionizing source, it is difficult to accurately deal with the large parameter space governing photoevaporation.

In MBH06, we attempted to tackle this issue with a statistical approach. By giving up on simulating ``realistic" (as much as this is possible given present uncertainties) relic HII regions left behind by deceased Pop III stars, we were able to explore larger swaths of parameter space and have a fairly large sample of second-generation objects.  We did this by applying various uniform ultraviolet (UV) and LW backgrounds on our entire simulation box, and statistically comparing the resulting evolution against the fiducial run without radiation.  Although this approach clearly does not realistically treat the photoionization around a particular source, it might provide a good statistical description of a ensemble of reionization studies.  Clearly, at the onset of early reionization not all halos lie within HII regions, as they do in our models. Our primary purpose is not to simulate early reionization or the photoevaporation of well-formed Pop III halos in detail (which would require radiative transfer) but to evaluate the impact of relic HII regions on the formation of halos at later times. Our uniformly illuminated boxes provide an ensemble of evaporated halos at high redshift that statistically sample the effects of early photoionization on the assembly of halos and formation of cold dense gas at lower redshifts.

However, since our analysis in MBH06 focused on a highly biased region which went non-linear at a high redshift, we could not follow the evolution to redshifts much lower than $z\sim20$. This meant that we could only extrapolate the observed trends suggesting that the feedback was transient.  \citet{OH03} claim that, even in the presence of a weak LWB, excess entropy will eventually suppress star formation in protogalaxies forming inside relic HII regions. In MBH06, we were unable to conclusively verify or refute this claim.

In this work, we extend the study by MBH06 by analyzing radiative feedback in a less-biased region, whose typical proto-galaxies form at lower redshifts, $z\sim$ 13--20.  Given that the 5-yr {\it WMAP} polarization data place a constraint on the reionization redshift (assuming instantaneous reionization) of $z=11.0 \pm 1.4$ \citep{Dunkley08}, such cosmological regions are likely to host the majority of feedback effects from Pop III stars, before a persistent UVB and/or a strong LWB become entrenched (e.g. \citealt{HRL96}).  We apply the same statistical approach as in MBH06, and study various combinations of transient UVBs and persistent LWBs.  As such, we attempt to provide a framework for analytically incorporating such radiative feedback into semi-analytic, and large-scale numerical and semi-numerical studies.

In \S \ref{sec:sims} we enumerate and describe the simulations which are used in this work. In \S \ref{sec:trans}, we study the initial radiative feedback caused by a transient UVB, adding a persistent LWB in \S \ref{sec:LW}.  Then in \S \ref{sec:pos}, we describe a related mechanism that results in eventual positive feedback, regardless of the strength of the UVB. In \S \ref{sec:ss}, we discuss the impact of neglecting self-shielding.  Finally, we offer our conclusions in \S \ref{sec:conc}.

Throughout this paper, we adopt the background cosmological parameters
($\Omega_\Lambda$, $\Omega_{\rm M}$, $\Omega_b$, n, $\sigma_8$, $H_0$)
= (0.7, 0.3, 0.047, 1, 0.92, 70 km s$^{-1}$ Mpc$^{-1}$), consistent
with the measurements of the power spectrum of CMB temperature
anisotropies by the first year of data from the {\it WMAP} satellite
\citep{Spergel03}.  Although the 5-yr data prefers a slightly lower value of $\sigma_8=0.82$ \citep{Dunkley08, Komatsu08} which somewhat delays structure formation, we keep the cosmology the same as in MBH06, to facilitate direct comparison.  Unless stated otherwise, we quote all quantities in comoving units.

\section{Simulations}
\label{sec:sims}

We use the Eulerian adaptive mesh refinement (AMR) code Enzo, which is
described in greater detail elsewhere \citep{Bryan99, NB99}.  Our
simulation volume is 1 $(\hMpc)^3$, initialized at $\zinit=99$ with
density perturbations drawn from the \citet{EH99} power spectrum.  Our root grid is
$128^3$. We have two additional static levels of refinement inside
a central 0.25 $\hMpc$ cube.  In addition, grid
cells inside the central region are allowed to dynamically refine so
that the Jeans length is resolved by at least 4 grid zones and no grid
cell contains more than 4 times the initial gas mass element.  Each
additional grid level refines the mesh length of the parent grid cell
by a factor of 2.  We allow for a maximum of 10 levels of refinement
inside the refined central region, granting us a spatial resolution of
7.63 $h^{-1}~{\rm pc}$.  This comoving resolution translates to 0.36
$h^{-1}~{\rm proper~pc}$ at $z=20$.

As stated above, our simulation runs were set up to facilitate comparison with MBH06.  Users interested in details of the simulations are encouraged to consult MBH06 and \citet{MBA01}.
There is one notable difference between our runs here and those in MBH06: our central refined region for these runs is {\it not} centered on the highest density region of the box.  Instead is it chosen to be more typical of the regions expected to host the bulk of halos which form prior to any significant cosmological reionization.  Specifically, the density inside our central $(0.25 \hMpc)^3$ corresponds to a 0.75 $\sigma$ mass fluctuation of an equivalent spherical volume (in MBH06, we studied a 2.4 $\sigma$ region).  The less overdense region allows us to extend the analysis of MBH06 to lower redshifts, and test the robustness of the conclusions of MBH06 on a less-biased and more typical region.

As shown in Table \ref{tbl:runs}, we have performed four different
runs without a LW background, distinguished by the duration or
amplitude of the assumed UVB, and
eight additional runs that include an additional constant LW background.  Again, these runs are analogous runs to those in MBH06.
For the UV radiation we assume an isotropic background flux with a
$T=2\times10^4$K blackbody spectral shape, normalized at the hydrogen
ionization frequency, $h \nu_H$ = 13.6 eV.  This spectrum is softer than 
the $\sim 8 \times 10^4$ K blackbody spectrum predicted to be produced 
by Pop III stars (e.g. \citealt{Schaerer02}), which leads us to find a lower 
reionization temperature (by about a factor of two) than we would have
found with a harder spectrum.  This results in outflow velocities (and therefore halo evacuation times)
which are also too low by this factor.  

The values of $\Juvb$ 
are shown in Table \ref{tbl:runs} in units of $10^{-21}{\rm
ergs~s^{-1}~cm^{-2}~Hz^{-1}~sr^{-1}}$.  The \nothing\ run contains no
UV radiation, and serves mainly as a reference run. The \heatinglow\
and \heatinghigh\ runs include a UVB with $\Juvb = 0.08, 0.8$,
respectively.  The value of $\Juvb=0.08$ was chosen to correspond
to the mean UV flux expected inside a typical HII region surrounding a
primordial star (e.g., \citealt{ABS05}).  As we do not include
dynamically expanding HII regions in our code, the \heatinghigh\ and
\flash\ (in which the UVB is turned on for a negligibly short duration) runs can be viewed as extremes, corresponding to conditions
close to the center and close to the edge of the HII region,
respectively.  
More generally, studying a range of values of $J_{\rm
UV}$ is useful, since the radiative properties of high-redshift sources are uncertain.
In the latter two runs, the UVB is turned on at
$\zuvbon=25$ and turned off at $\zuvboff=24.64$. This redshift range
corresponds to a typical theoretical stellar lifetime, $\sim3$ Myr, of
a $\sim 100 \Msun$ primordial (Pop-III) star \citep{Schaerer02}.  The
flash ionization run, \flash, instantaneously sets the gas temperature
to T=15000 K and the hydrogen neutral fraction to $\nf = 10^{-3}$
throughout the simulation volume, but involves no heating thereafter.
Again, our adopted temperature is probably too low by a factor of two
compared to gas heated by Pop III stars.
  Finally, the eight runs in
the bottom half of the Table repeat pairs of the \nothing\ and
\heatinghigh\ runs with four different constant LW backgrounds
($\Jlwb=$ 10$^{-4}$, 0.001, 0.01, and 0.1, normalized at 12.87eV in units of
$10^{-21}{\rm ergs~s^{-1}~cm^{-2}~Hz^{-1}~sr^{-1}}$, and assumed to be
frequency--independent within the narrow LW band).

\begin{table}
\vspace{0.2cm}
\caption{Summary of Simulation Runs}
\vspace{-0.2cm}
\label{tbl:runs}
\begin{center}
\begin{tabular}{ccccc}
\hline
Run Name & $\Juvb$ &  $\zuvbon$ & $\zuvboff$ &   $\Jlwb$  \\
\hline
\hline
\multicolumn{5}{c}{Runs without a LW background}\\
\hline
\nothing      &   0   &    NA   &   NA    & 0 \\
\flash        &   --  &    25   &   25    & 0 \\
\heatinglow   &  0.08 &    25   &  24.62  & 0 \\
\heatinghigh  &  0.8  &    25   &  24.62  & 0 \\
\hline
\multicolumn{5}{c}{Runs with a LW background}\\
\hline
\LWfournothing      &   0   &    NA   &   NA    & $10^{-4}$\\
\LWfourheatinghigh  &  0.8  &    25   &  24.62  & $10^{-4}$\\
\LWthreenothing      &   0   &    NA   &   NA    & $10^{-3}$\\
\LWthreeheatinghigh  &  0.8  &    25   &  24.62  & $10^{-3}$\\
\LWtwonothing      &   0   &    NA   &   NA    & 0.01\\
\LWtwoheatinghigh  &  0.8  &    25   &  24.62  & 0.01\\
\LWonenothing      &   0   &    NA   &   NA    & 0.1\\
\LWoneheatinghigh  &  0.8  &    25   &  24.62  & 0.1\\
\hline
\end{tabular}\\
\end{center}
\end{table}

We obtain the dark matter (DM) and baryonic mass of halos, in the same manner as MBH06.  Similarly, we use the fraction of
total gas within the virial radius which is cold and dense (CD),
$\fcd$ as a proxy for the gas reservoir available to form stars.  By cold, we mean gas whose temperature is $<0.5 \Tvir$, where
$\Tvir$ is the halo's virial temperature (for how virial temperatures
are associated with halos in the simulation, see \citealt{MBA01}).  By
dense, we mean gas whose density is $> 10^{19}$ $\Msun$ Mpc$^{-3}$
$\approx$ 330 cm$^{-3}$, roughly corresponding to the density at which
the baryons become important to the gravitational potential in the core,
taken to be an immediate precursor to primordial star formation
\citep{ABN02}.

Additionally, we discount halos which have been substantially
contaminated by the large (low-resolution) DM particles originally outside of our
refined region which enter it as it becomes nonlinear.  Specifically, we remove from our analysis halos with an average DM particle mass greater than $\sim$140\% of the refined region's
DM mass resolution, $747\Msun$.  However, since there are small stocastic variations in the amount of contamination between the runs, we first apply this cut-off in the \nothing\ run, and then remove the same halos from the analysis of the other runs.  This way, we study the same halos in each run.

Another possible source of
contamination arises from closely separated halos.  If some CD gas
belonging to a halo is within another halo's virial radius (most
likely in the process of merging), the other halo could undeservedly
be flagged as containing CD gas as well.  To counteract this, we set
$\fcd=0$ for all halos without CD gas in their inner-most spherical shell.  CD gas invariably first forms close to the halo's core, where the cooling times are the shortest; thus halos which contain CD gas at large radii but none in their core are most likely in the process of merging with a halo with genuine CD gas. Note that this algorithm is different than the one used in MBH06, where we discriminated based on mass and separation between halos.  In this work such contamination was more prevalent and so we devised what proved to be a more robust means of discrimination.

Finally, we note that we do not include HD cooling in these simulations.  As is shown in \citet{McGreer08} and \citet{Yoshida07}, HD cooling is important in primordial halos only in the density range $10^2 - 10^6$ cm$^{-3}$, where H$_2$ can bring the temperature below about 200 K.  Therefore it does not change the amount of CD gas, our primary diagnostic (although it would lead to relatively small changes in the predicted masses of the resulting stars).

\section{Transient Feedback Following a Transient UVB}
\label{sec:trans}

\begin{figure*}
\vspace{+0\baselineskip}
{
\includegraphics[width=0.245\textwidth]{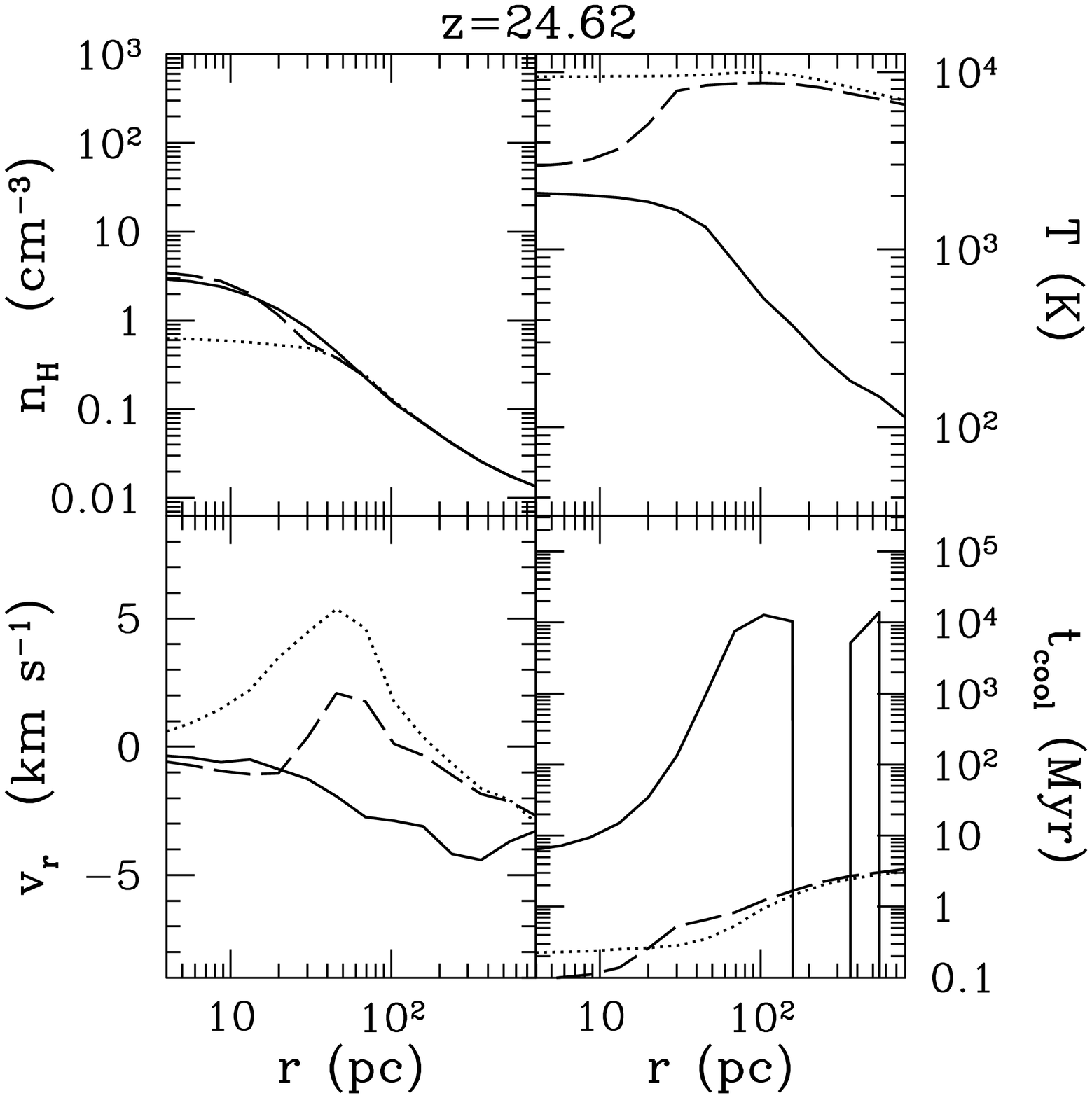}
\includegraphics[width=0.245\textwidth]{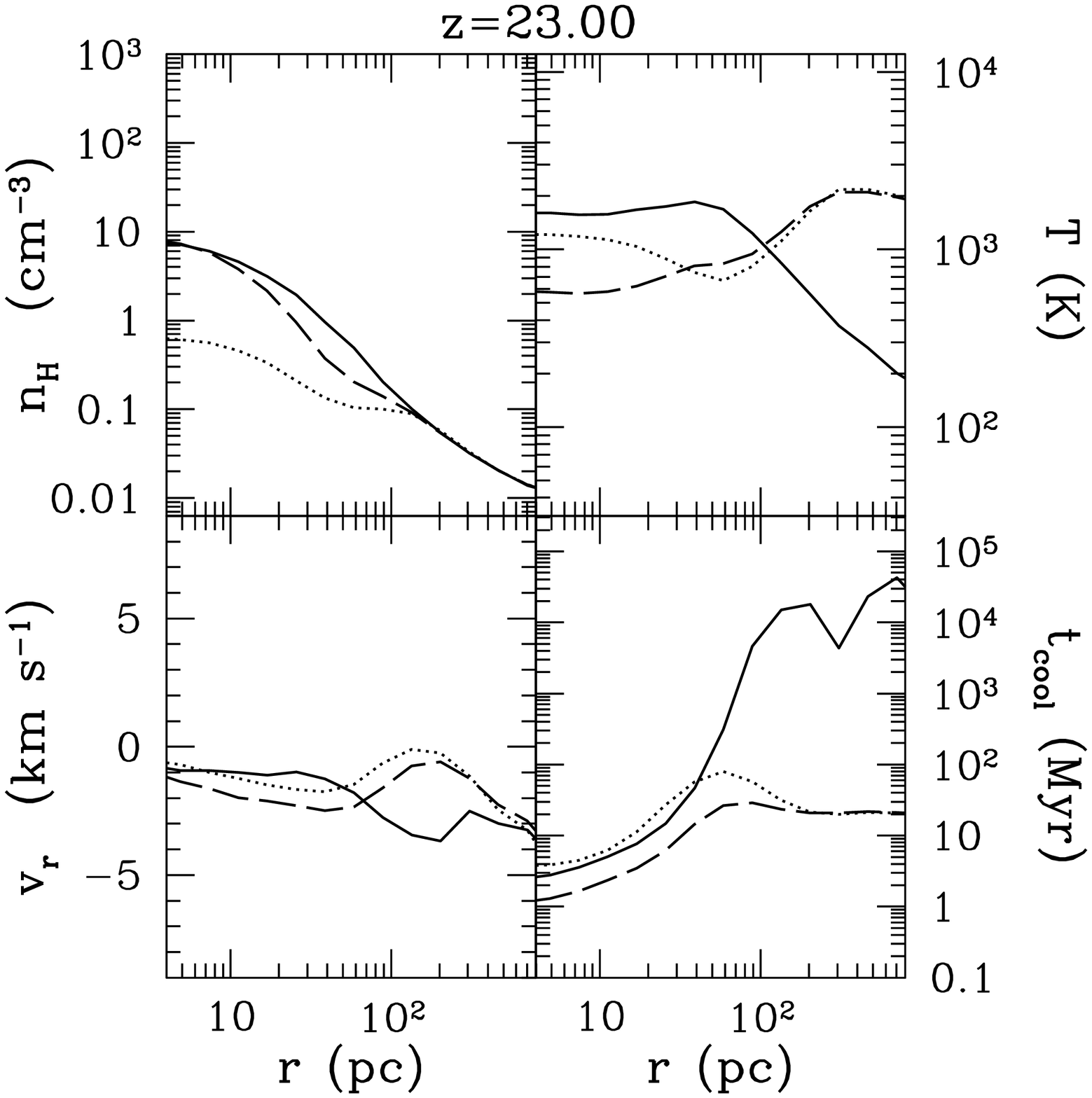}
\includegraphics[width=0.245\textwidth]{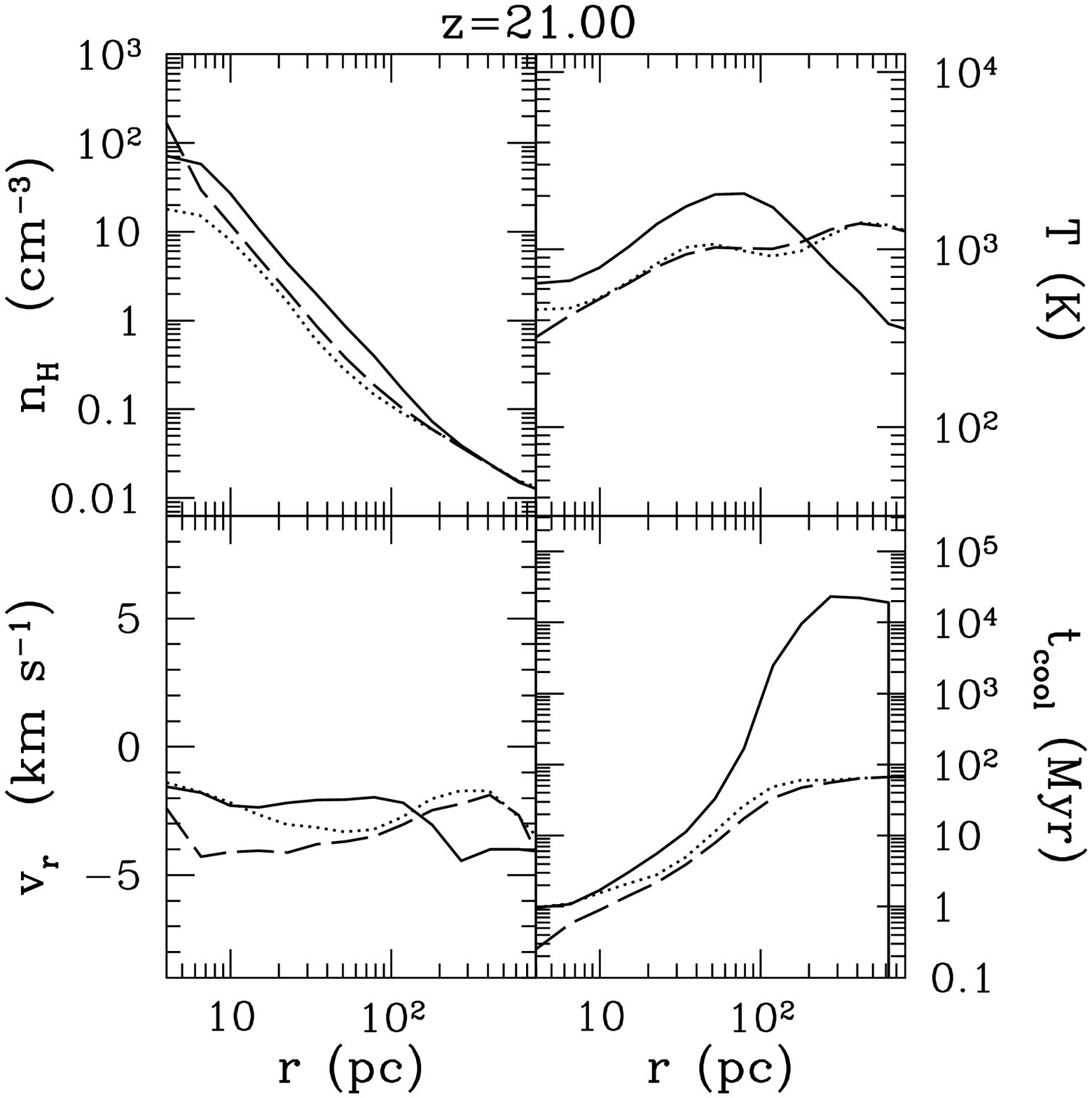}
\includegraphics[width=0.245\textwidth]{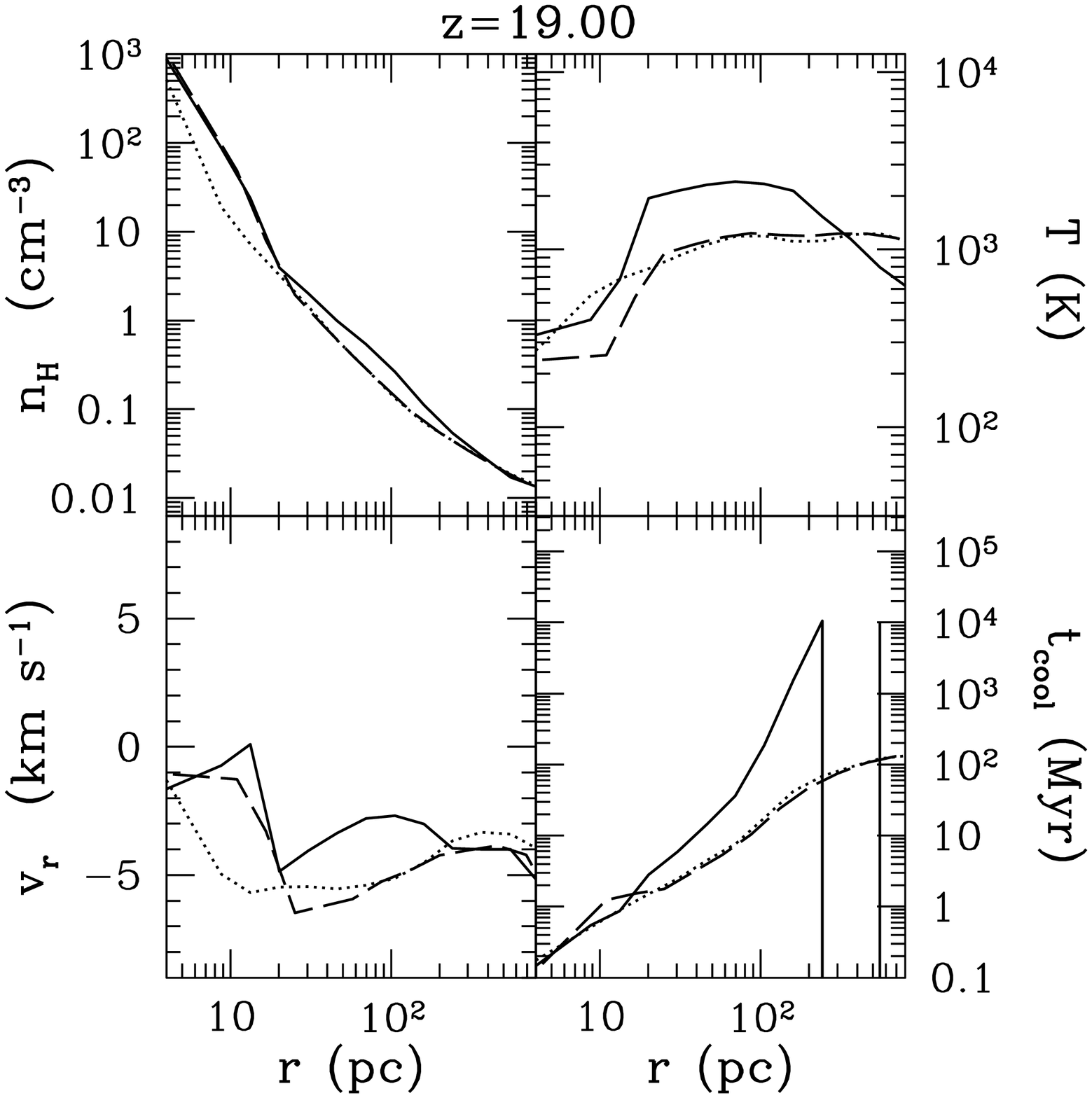}
\vskip0.0pt
}
{
\includegraphics[width=0.245\textwidth]{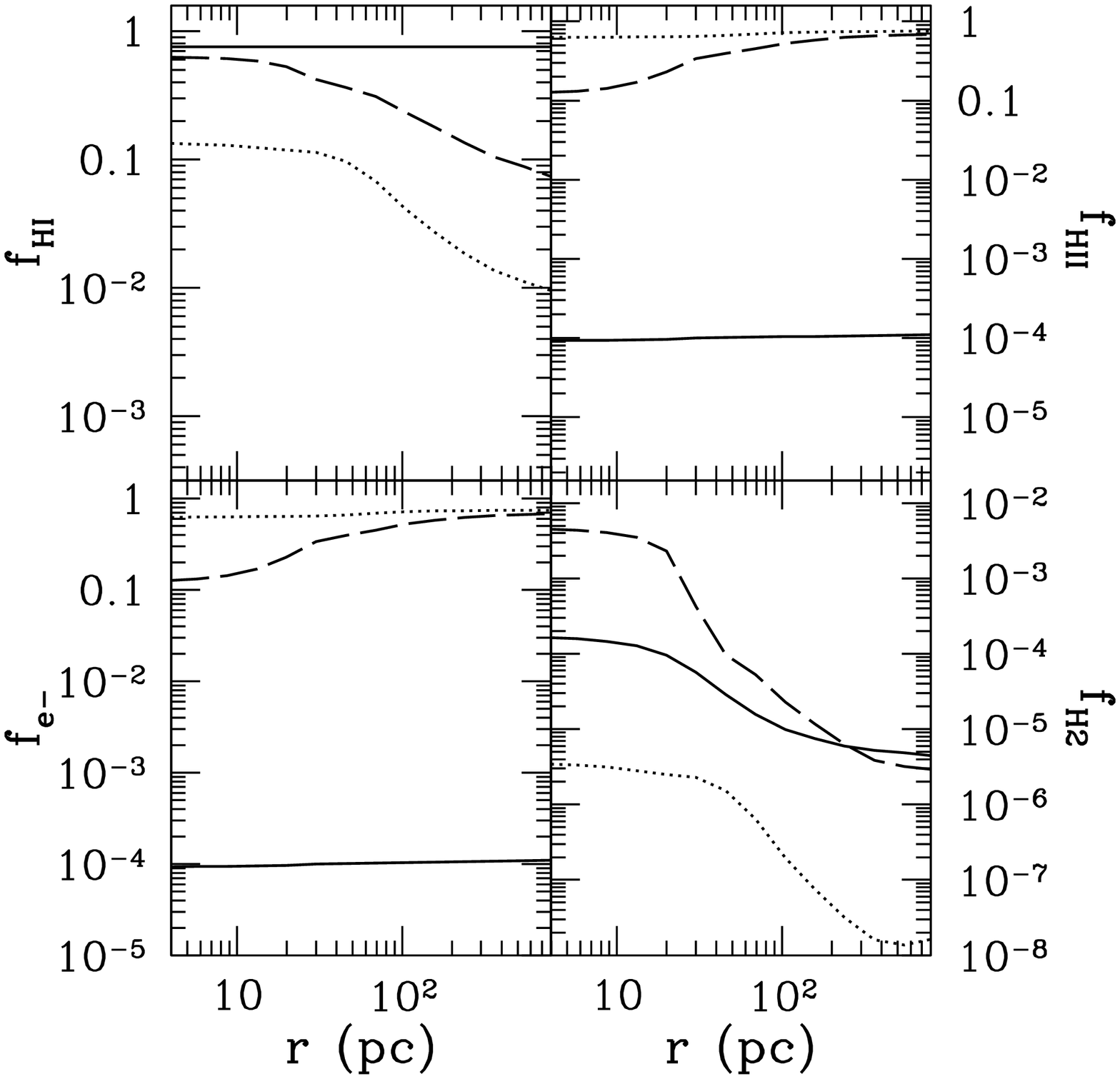}
\includegraphics[width=0.245\textwidth]{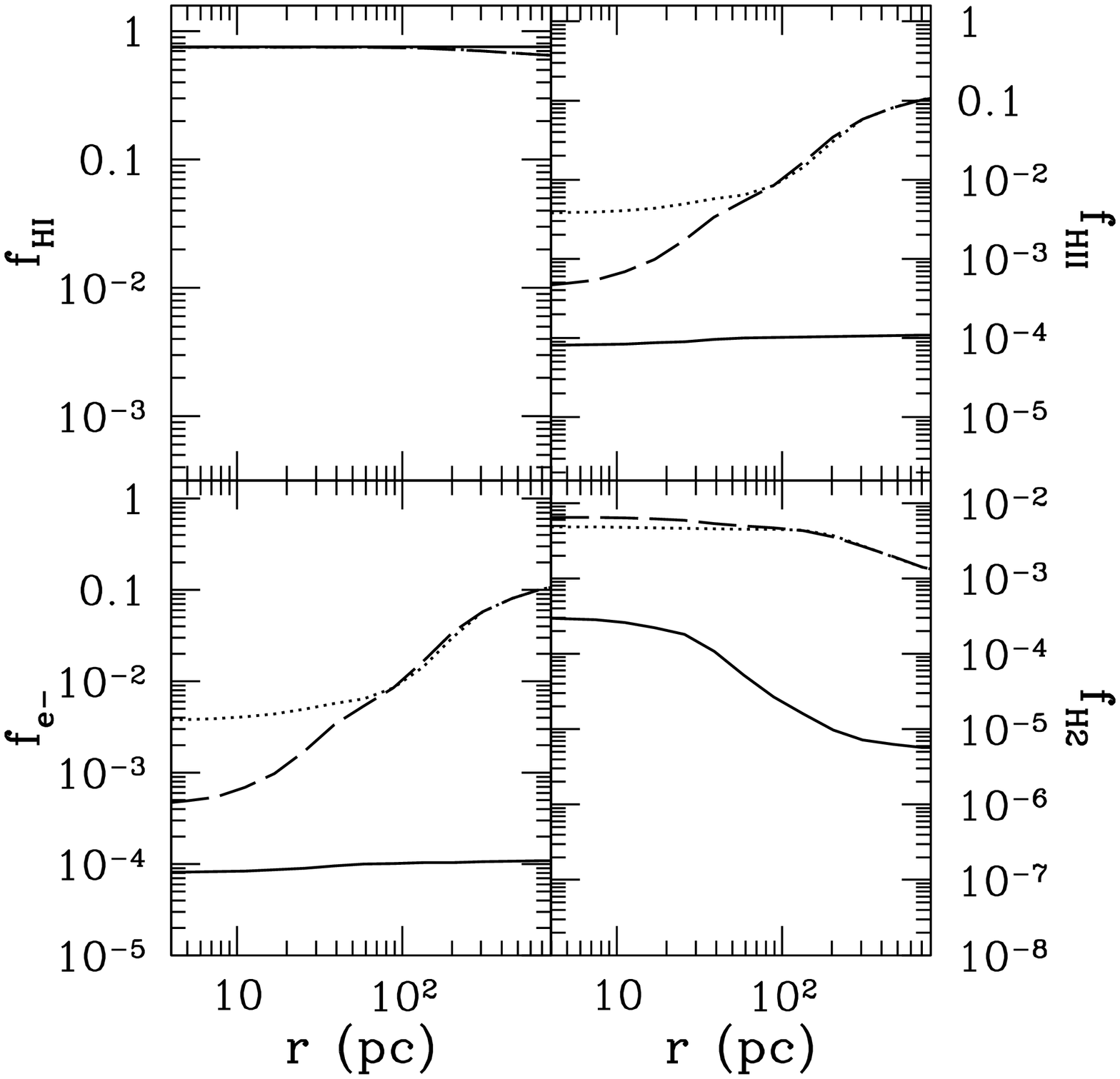}
\includegraphics[width=0.245\textwidth]{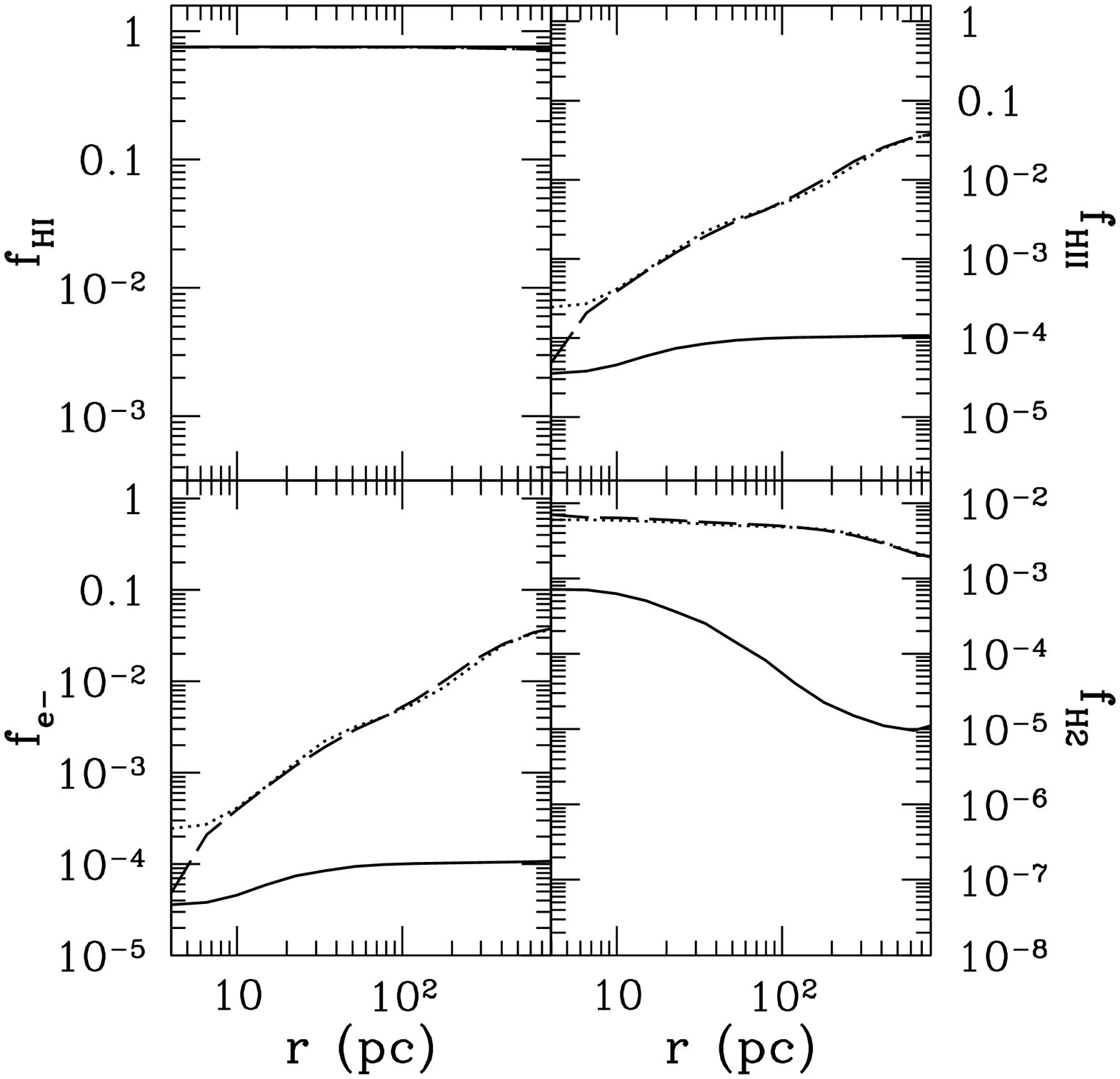}
\includegraphics[width=0.245\textwidth]{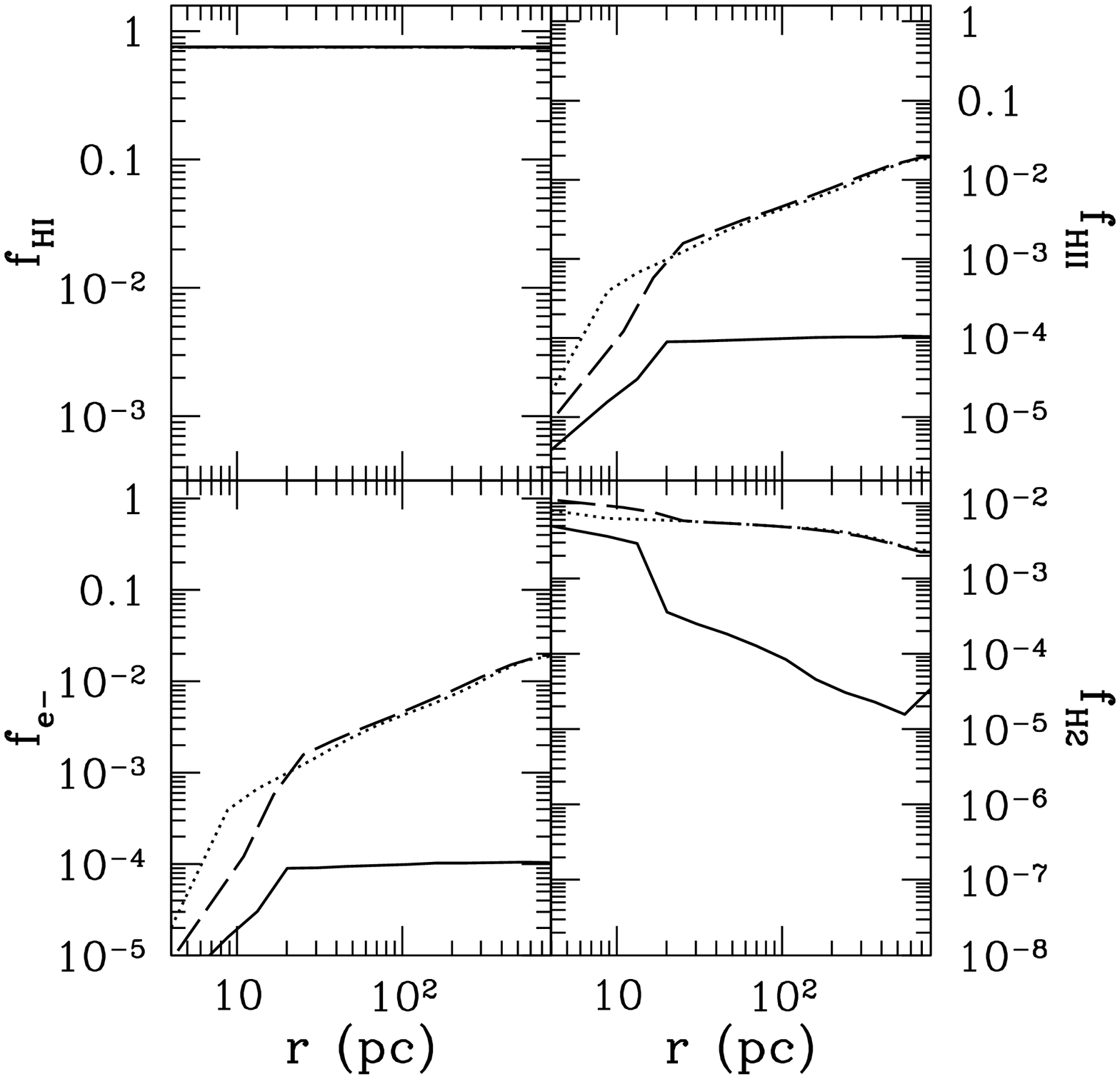}
}
 \caption{
Spherically averaged radial
profiles of the same halo in the \nothing\ ({\it solid lines}), \heatinglow\ ({\it dashed lines}) and
\heatinghigh\ ({\it dotted lines}) simulation runs.  This halo was first able to form CD gas at $z=$ 20, 21, 19 in the \nothing, \heatinglow, and \heatinghigh\ runs, respectively.
The top and bottom pairs of figures show snapshots at $z =$ 24.62, 23, 21, 19 ({\it left to right}).
The halo mass in the \nothing\ run increases from M=$4.2\times10^5 \Msun$ at $z=24.62$, to M=$1.4\times10^6 \Msun$ at $z=19$.
All quantities are shown in proper (not comoving) units.
The {\it Upper panels} show the hydrogen density, mass-weighted gas
temperature, gas cooling time, and radial velocity ({\it clockwise
from upper left}).  The {\it Bottom panels} show mass fractions of HI,
HII, H$_2$, and the number fraction of $e^-$ ({\it clockwise from
upper left}).
\label{fig:early_profiles}
}
\vspace{-1\baselineskip}
\end{figure*}

We first wish to confirm whether our conclusions in MBH06 also apply to this new sample of halos, which form at later times in a less overdense region of the universe.  In MBH06, we found the following.  The UVB ionizes and heats the gas in our simulation box.\footnote{It is important to note that we are operating in the optically thin limit; the accuracy and impact of that approximation is discussed in section~\ref{sec:ss}.}  Gas inside halos which used to be at or close to hydrostatic equilibrium feels an increased pressure gradient from the rise in temperature, resulting in an outward--moving pressure shock.  Once the UVB is turned off, the gas quickly cools to $T\sim10^3$ K through a combination of atomic, molecular hydrogen and Compton cooling. The molecular hydrogen formation time is shorter than the recombination time and a large amount of molecular hydrogen is produced, $x_{\rm H_2} \sim$ few $\times 10^{-3}$, irrespective of the density and temperature (see also \citealt{OH02}).\footnote{\citealt{Glover08} have recently shown that including cooling due to excitations of $H_2$ through collisions with protons and electrons can lead to a lower molecular hydrogen fraction by a factor of 2.  This would result in a longer delay in star-formation than shown here.}  The pressure shock starts to dissipate and gas expelled from the center of the halo starts falling back in (for a more complete description, readers are encouraged to see MBH06).

Quantitatively, we concluded that the net impact of a transient UVB on star-formation could be understood simply in terms of the molecular hydrogen cooling time-scale:

\begin{align}
\label{eq:t_H2}
\tHtwo &= \frac{1.5 k_B T}{\Lambda_{\rm H_2}} \frac{n_g}{n_{\rm HI} n_{\rm H_2}}\\
&\approx  \frac{1.5 k_B T}{\Lambda_{\rm H_2}} \frac{1}{n_g x_{\rm H_2}}\\
&\approx 4 \left( \frac{T}{10^3 K} \right)^{-2.5}
                   \left( \frac{x_{\rm H_2}}{3 \times 10^{-3}} \right)^{-1}
                   \left( \frac{n_g}{1 \rm cm^{-3}} \right)^{-1} {\rm Myr}~ , \nonumber
\end{align}

\noindent Here, $k_B$ is the Boltzmann constant, $T$ is the temperature,
$\Lambda_{\rm H_2}$ is the H$_2$ cooling function, $x_e$ is the free
electron number fraction, $n_g$, $n_{\rm HI}$, and $n_{\rm H_2}$ are
the number densities of all baryons and electrons, neutral hydrogen,
and H$_2$, respectively, and the second equality is accurate shortly after $\zuvboff$, since the recombination times at high redshifts and inside halos are very short.

We found that the net {\it delay} in star-formation caused by the transient UVB is simply the ratio of the H$_2$ cooling time in the run with the UVB, to the H$_2$ cooling time in the run without UV heating.  This ratio is calculated near the halo core, shortly after the UVB disappears (when the shock first starts to dissipate and the gas behind it starts infalling again).  For example, the delay in star-formation of a halo in the \heatinghigh\ run with respect to the same halo in the \nothing\ run is:

\begin{equation}
\label{eq:delay}
\fdelay \sim
\frac{\tHtwo^{\rm\heatinghigh}}{\tHtwo^{\rm\nothing}} \sim
\left( \frac{T^{\rm\nothing}}{T^{\rm\heatinghigh}} \right)^{2.5}
\left( \frac{n^{\rm\nothing}_g}{n^{\rm\heatinghigh}_g} \right)
\left( \frac{x^{\rm\heatinghigh}_{\rm H_2}}{x^{\rm\nothing}_{\rm  
H_2}} \right)^{-1} ~.
\end{equation}

We will now demonstrate that eq. (\ref{eq:delay}) is equally adept at quantifying feedback in our new simulations.  First, we show the spherically averaged radial profiles of the same (typical) halo in the \nothing\ ({\it solid lines}), \heatinglow\ ({\it dashed lines}) and \heatinghigh\ ({\it dotted lines}) simulation runs in our Figure \ref{fig:early_profiles}.  This halo, which has a mass of $4\times10^5 \Msun$ at $z=24.62$, was first able to form CD gas at $z=$ 20, 21, 19 in the \nothing, \heatinglow, and \heatinghigh\ runs, respectively. The top and bottom pairs of figures are snapshots at $z =$ 24.62, 23, 21, and 19 ({\it left to right}). The {\it upper panels} show the hydrogen density, mass-weighted gas temperature, gas cooling time, and radial velocity ({\it clockwise from upper left}).  The {\it bottom panels} show mass fractions of HI, HII, H$_2$, and the number fraction of $e^-$ ({\it clockwise from upper left}).

The profiles confirm the qualitative story line from the beginning of this section. The outward-moving pressure shock is evident in both UVB runs from the density and velocity panels at $z=\zuvboff$, just before the UVB was turned off.  This gas has already cooled to $T=10^3$ K and started to collapse back onto the halo only a few Myr afterward at $z=23$.  Note that the outflow and the resulting suppression of gas near the core is stronger in the run with a stronger UVB.  Halos in the runs with a UVB also experience a strong boost in the H$_2$ fraction following $\zuvboff$, with the increase being insensitive to the strength of the UVB.  Formation of CD gas in this halo is notably delayed until $z=19$ in the \heatinghigh\ run, while being mildly expedited in the \heatinglow\ run.  Can this behavior be understood through eq. (\ref{eq:delay})?

Let us focus on the output we have shortly after $\zuvboff$, at $z=23$.  We can note that at this redshift, the pressure-driven shock has almost dissipated and reached zero velocity with gas recommencing infall behind the shock.  Plugging in appropriate values into eq. \ref{eq:delay} from the inner annulus of the halo, we obtain values of $\fdelay\approx$ 0.5 and 1.5 for the \heatinglow\ and \heatinghigh\ runs, in good agreement with the simulation outcomes.  This behavior again supports our conclusions in MBH06, where the same analysis held for a different, randomly-chosen halo, and implies that we understand the physical picture associated with feedback in relic HII regions.

In order to quantify the suppression of CD gas by a transient UVB on our entire sample of halos, 
 we define (as in MBH06) the cumulative, fractional suppression of the number of newly formed halos as
\begin{equation}
\label{eq:delN}
\delN \equiv \frac{\Nrun(z) - \Nrun(\zuvbon)}{N_{\rm cd}^{\rm
\nothing}(z) - N_{\rm cd}^{\rm \nothing}(\zuvbon)} - 1 ~ ,
\end{equation}
where $N_{\rm cd}^{\rm \nothing}(z)$ and $\Nrun(z)$ are the total
number of halos with CD gas at redshift $z$ in the \nothing\ run and
some given run $i$, respectively.  This expression is well--defined
for $N_{\rm cd}^{\rm \nothing}(z)$ $>$ $N_{\rm cd}^{\rm
\nothing}(\zuvbon)$; for $N_{\rm cd}^{\rm \nothing}(z)$ = $N_{\rm
cd}^{\rm \nothing}(\zuvbon)$, we set $\delN \equiv 0$.  Note that by
definition, $N_{\rm cd}^{\rm \nothing}(\zuvbon)$ = $\Nrun(\zuvbon)$.

Similarly, we define the cumulative, fractional suppression of the CD
gas mass as
\begin{equation}
\label{eq:delM}
\delM \equiv \frac{\Mrun(z) - \Mrun(\zuvbon)}{M_{\rm cd}^{\rm
\nothing}(z) - M_{\rm cd}^{\rm \nothing}(\zuvbon)} - 1 ~ ,
\end{equation}
where $M_{\rm cd}^{\rm \nothing}(z)$ and $\Mrun(z)$ are the total mass
of CD gas at redshift $z$ in the \nothing\ run and some given run $i$,
respectively.  The total CD gas mass is obtained by merely summing the
CD gas masses for all of the halos in the simulation.  As for
equation~(\ref{eq:delN}), this expression is well--defined for $M_{\rm
cd}^{\rm \nothing}(z)$ $>$ $M_{\rm cd}^{\rm \nothing}(\zuvbon)$, and
for $M_{\rm cd}^{\rm \nothing}(z)$ = $M_{\rm cd}^{\rm
\nothing}(\zuvbon)$, we set $\delM \equiv 0$.

\begin{figure}
\myputfigure{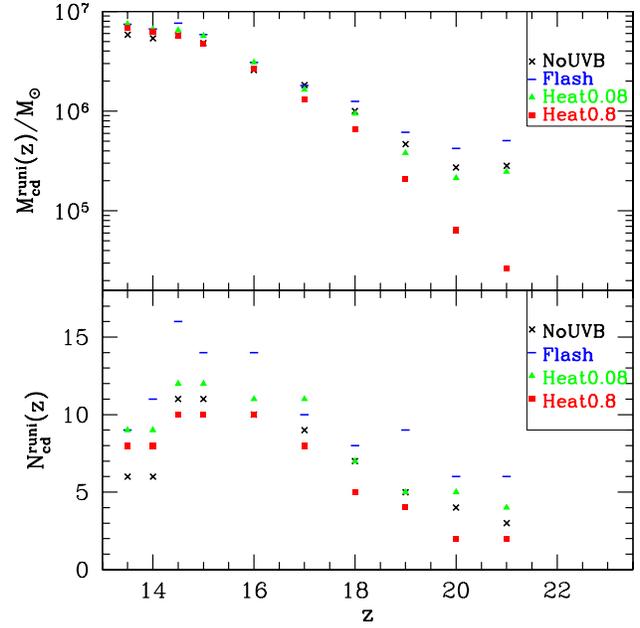}{3.3}{0.5}{.}{0.}  \caption{Values of
$\Mrun(z)$ ({\it top panel}) and $\Nrun(z)$ ({\it bottom panel}) as
defined in equations (\ref{eq:delM}) and (\ref{eq:delN}).  The results
are displayed for the \nothing\ ({\it crosses}), \flash\ ({\it
dashes}), \heatinglow\ ({\it triangles}), and \heatinghigh\ ({\it
squares}) simulation runs.
\label{fig:NM}}
\vspace{-1\baselineskip}
\end{figure}

Equations (\ref{eq:delN}) and (\ref{eq:delM}) provide an estimate of
how the CD gas has been affected by the presence of a UVB, {\it
following} the turn-on redshift of the UVB, $\zuvbon$ (the values at
$\zuvbon$ are subtracted in order to provide a more sensitive measure
of {\it relative} changes of CD gas).  As defined above, $\delN = 0$
and $\delM = 0$ if the UVB has no effect.  If the effect of a UVB is
positive, resulting in positive feedback, $\delN$ and $\delM$ would be
positive.  If the effect of the UVB is negative, $\delN$ and $\delM$
would be negative.

In Figure \ref{fig:NM}, we plot the values of $\Mrun(z)$ ({\it top
panel}) and $\Nrun(z)$ ({\it bottom panel}) in our four
simulation runs with no LW background: ${\rm run}i$ = \nothing\ ({\it crosses}), \flash\
({\it dashes}), \heatinglow\ ({\it triangles}), and \heatinghigh\
({\it squares}).  We note that the number of halos with CD gas increases in all runs until $z\sim15$, when our central refined regions starts becoming more non-linear.  At these lower redshifts, halos become contaminated with large DM particles entering from outside the refined region and we remove them from our analysis sample (see \S \ref{sec:sims}).

\begin{figure}
\myputfigure{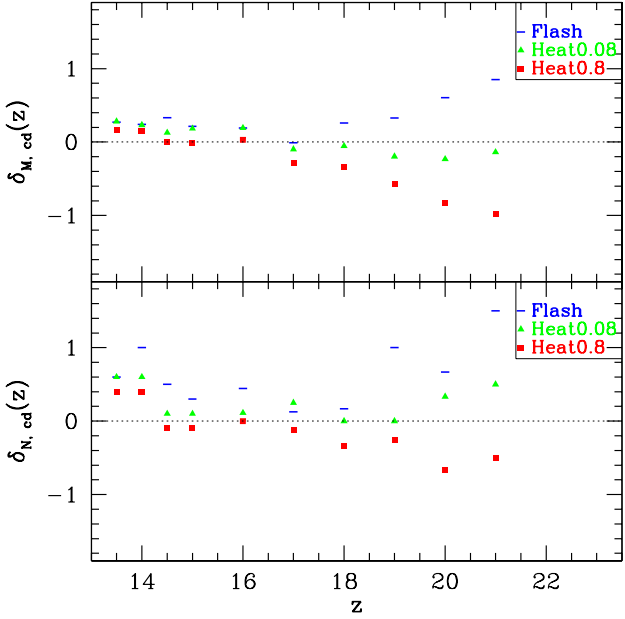}{3.3}{0.5}{.}{0.}  \caption{
Values of $\delM$ ({\it top panel}) and $\delN$ ({\it bottom panel}) as defined
in equations (\ref{eq:delM}) and (\ref{eq:delN}).  The results are
derived from Figure~\ref{fig:NM} and displayed for the \flash\ ({\it
dashes}), \heatinglow\ ({\it triangles}), and \heatinghigh\ ({\it
squares}) simulation runs.
\label{fig:delta}}
\vspace{-1\baselineskip}
\end{figure}

The corresponding values of $\delM$ and $\delN$ are
plotted in Figure \ref{fig:delta} in the top and bottom panels,
respectively. Although some of the notable fractional
changes shown in Figure \ref{fig:delta} might appear statistically
insignificant due to the small number statistics inferred from Figure
\ref{fig:NM}, it should be noted that these runs are not uncorrelated
experiments.  In other words, each of our runs in Table \ref{tbl:runs}
is seeded with the same initial conditions, and so small relative
changes compared to the \nothing\ run are significant (i.e. the errors
are not Poisson).  Nevertheless, there are some minor differences between the runs in the amount of ``contamination'' by DM particles outside of our refined region. 

In MBH06, we presented two main inferences regarding the effects of a transient UVB (aside from the delay due to cooling times discussed above): (1) a ``critical'' background value of $\Juvb \sim 0.1$ separated positive from negative feedback (stronger backgrounds result in negative feedback, while weaker backgrounds result in positive feedback; see the discussion surrounding eq. \ref{eq:delay} above); (2) the feedback effects, regardless of the sign, were transient and started to ``disappear'' by $z\sim18$. However, we were unable to evolve our simulations below $z<18$, since the central refined region became too non-linear and further progress was computationally prohibitive.  Thus point (2) was an inference based on a figure analogous to Fig. \ref{fig:delta}. 

In this work we present results from a less biased region and are able to go to lower redshifts, $z\sim 13.5$.  Thus we can test both of the conclusions above and see if they are sensitive to the large-scale halo environment and redshift.  From figure \ref{fig:delta}, we see that {\it both} conclusions are valid for these simulations as well.  In other words, our transient (corresponding to the expected lifetime of a Pop III star) UVB results in transient feedback which disappears after $\sim30$\% of the Hubble time, with $\Juvb\sim0.1$ separating net positive and negative feedback. {\it Thus, these conclusions seem to be fairly insensitive to the large-scale overdensity of the simulated region, and by extension to the redshift-dependent properties of halos at $\zuvbon$.}

It is important to note a new result evident in Fig. \ref{fig:delta}: {\it all runs with a transient UVB experience eventual positive feedback at $z\lsim15$}. This is true regardless of the strength of the UVB.  We will examine this interesting result further in \S \ref{sec:pos}.

\subsection{Adding a Persistent LW Background}
\label{sec:LW}

Next, we investigate how the combination of transient UV heating and a persistent LW background impact star formation.  This is a more physically-relevant scenario, since a LW background is likely established before a large fraction of the Universe has been reionized \citep{HAR00}.

Photons in the LW band dissociate H$_2$ molecules; thus by definition they provide additional negative feedback, undercutting some of the enhancement in the H$_2$ abundances in relic HII regions.  Negative feedback from a LW background kicks in when the H$_2$ dissociation timescale becomes shorter than the H$_2$ formation timescale.  Since the formation timescale is inversely proportional to gas density, whereas the dissociation timescale is independent of density, the density decrease caused by UV heating should make halos more susceptible to the negative feedback of a LW background \citep{OH03} \footnote{Note that the temperature increase associated with a UVB does not contribute much this combined negative feedback because the ionized gas is able to Compton cool down to the virial temperatures of these minihalos.}.  We investigate these processes further below.

\begin{figure}
\vspace{+0\baselineskip}
\myputfigure{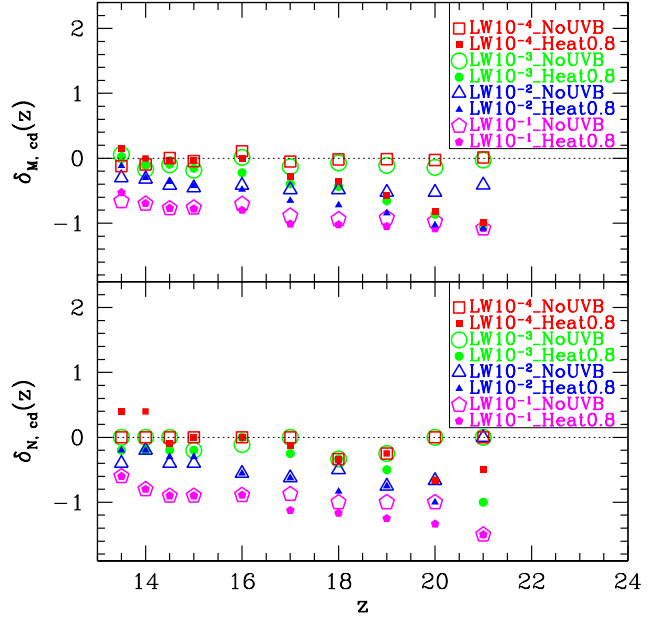}{3.3}{0.5}{.}{0.}
\caption{
Suppression of cold dense gas in halos in simulations runs that include a
persistent LW background. The LW background had specific intensities
of $J_{\rm LW}=10^{-4}, 10^{-3}, 0.01$, or $0.1$ (normalized at 12.87 eV, in
units of $10^ {-21}{\rm ergs~s^{-1}~cm^{-2}~Hz^{-1}~sr^{-1}}$). Each
of these three LW backgrounds is applied to both our \nothing\ and
\heatinghigh\ runs at $z=24.62$, and is subsequently left on.  Values
of $\delM$ ({\it top panel}) and $\delN$ ({\it bottom panel}) are
shown, as defined in equations (\ref{eq:delM}) and (\ref{eq:delN}).
\label{fig:LW_delta}}
\vspace{-1\baselineskip}
\end{figure}

Specifically, in Fig. \ref{fig:LW_delta}, we plot values of $\delM$ ({\it top panel}) and $\delN$ ({\it bottom panel}) for our runs including a LW background.  The LW background turns on at $z=24.62$ and remains on.  At the simplest level, we can see the suppression of molecular hydrogen (and hence cooling) due to the LW background by the more negative values of both $\delM$ and $\delN$ at fixed redshift.

Does this panel show evidence for the {\it additional} negative impact of the transient UVB discussed above?  Without UV heating, a LW background with a specific intensity of $\Jlwb=0.01$ is needed for notable negative feedback (see the blue empty triangles at $z\leq14$ in the lower panel).  This value is similar to the one found in MBH06, where we showed that by equating ${\rm H_2}$--cooling and ${\rm H_2}$--photodissociation one expects this critical LW background to scale as $\Jlwb \propto n_g f_{\rm H2}/T$.\footnote{Since the mean density of gas collapsing into halos in these simulations is less than a factor of two smaller than those in MBH06, we roughly expect this critical value of $\Jlwb$ to also be less than a factor of two smaller (at least in the linear regime).}
  When the UV heating is added, this critical value of $\Jlwb$ causing negative feedback decreases by a factor of $\sim10$ to $\Jlwb\sim10^{-3}$ (see the green solid circles in lower panel at $z\leq14$).  This decrease seems to confirm the above arguments.  However, such an interpretation is too simplistic and the negative impact of the LWB can be ameliorated by positive feedback inside the HII region at lower redshifts, as we shall see in \S \ref{sec:pos}.

From Fig. \ref{fig:LW_delta} one can also note that a value of $\Jlwb\sim 10^{-3}$--$10^{-2}$ separates feedback regimes dominated by a LW background from those dominated by our transient UVB.   This can be seen by the fact that the amount of suppression differs between the NoUVB and Heat0.8 cases at low values of the LW background (i.e. for $\Jlwb < 10^{-3}$), while at large values of $\Jlwb$, the amount of suppression is independent of the UVB.  Near the threshold value of $\Jlwb$, negative feedback transitions from being UV heating dominated (at $z \gsim 17$, or $\lsim 100$ Myr after $\zuvboff$) to being LW background dominated ($ z \lsim 17$, or $\gsim 100$ Myr after $\zuvboff$).  As highlighted in MBH06, this ``transition'' behavior can be understood as the
combined result of two effects: the UV heating is turned off, and its
impact is transient, while the critical LW
background scales roughly with the inverse of the density
\citep{HAR00, OH03} and hence a fixed LW background will have a larger
impact at lower densities or decreasing redshifts.

\section{Eventual Positive Feedback}
\label{sec:pos}

So far, we have not focused on the interesting new result hinted by the previous figures (e.g. see the $z\lsim14.5$ points in Fig. \ref{fig:delta}).  This figure shows evidence of delayed (approaching our lowest redshifts) {\it positive} feedback in {\it all} runs with a UVB.  This result is somewhat surprising given that the Heat0.8 run shows strong negative feedback initially after $\zuvboff$. In fact, as was already pointed out when discussing Fig. \ref{fig:delta}, the initial feedback, whether positive or negative, fades away in {\it all} runs by $z\sim17$.

\begin{figure*}
\vspace{+0\baselineskip}
{
\includegraphics[width=0.45\textwidth]{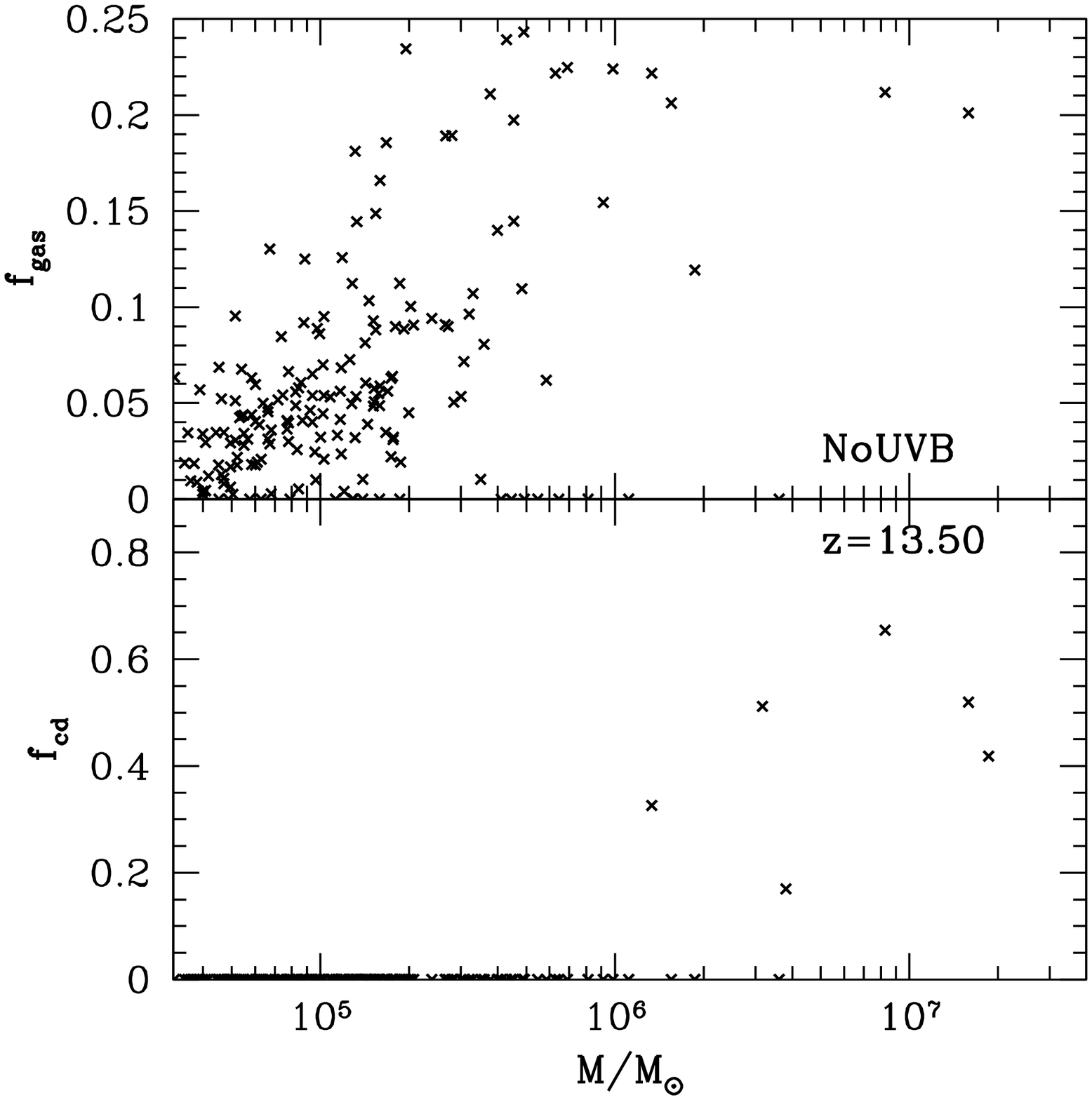}
\includegraphics[width=0.45\textwidth]{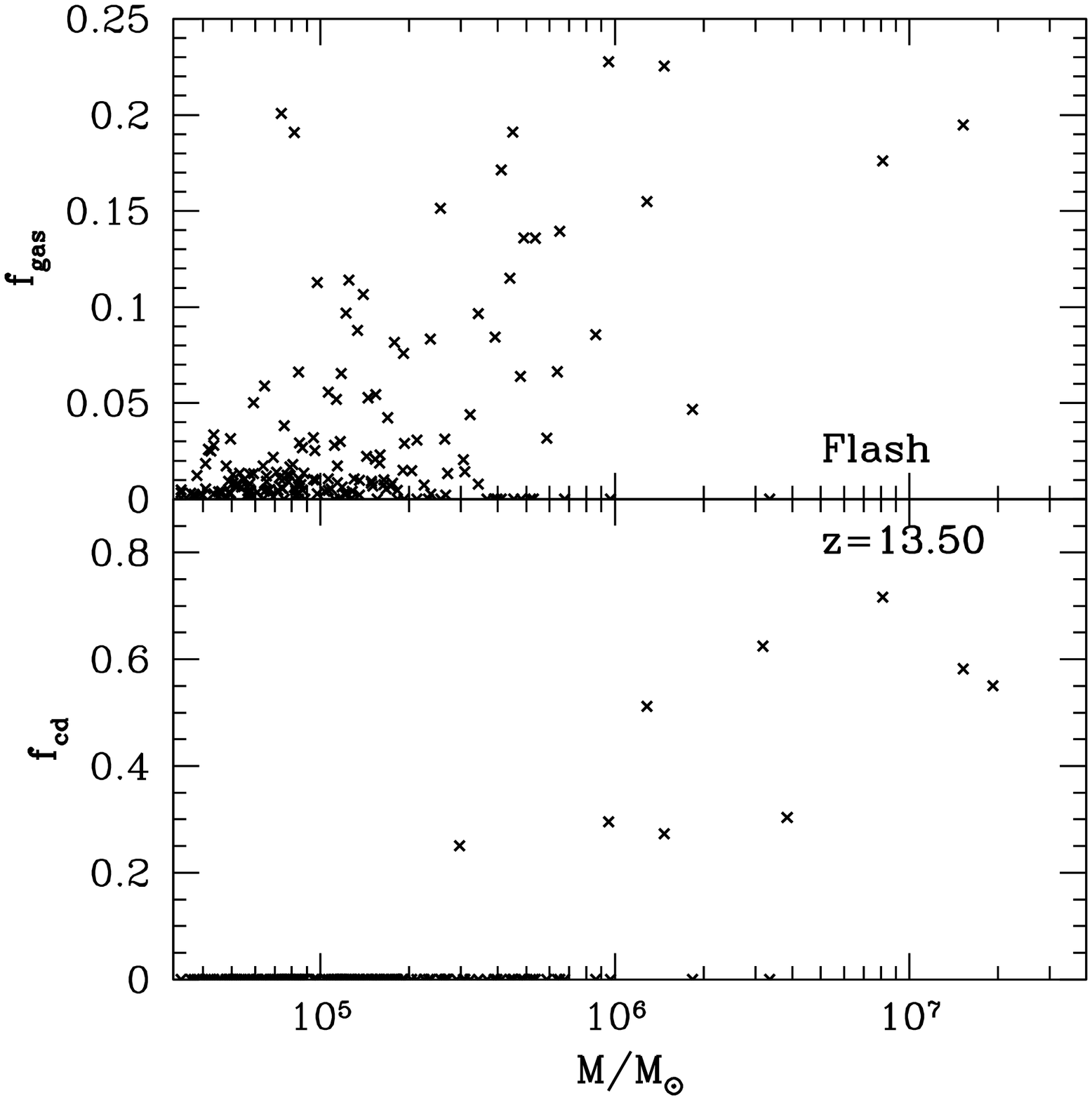}
\includegraphics[width=0.45\textwidth]{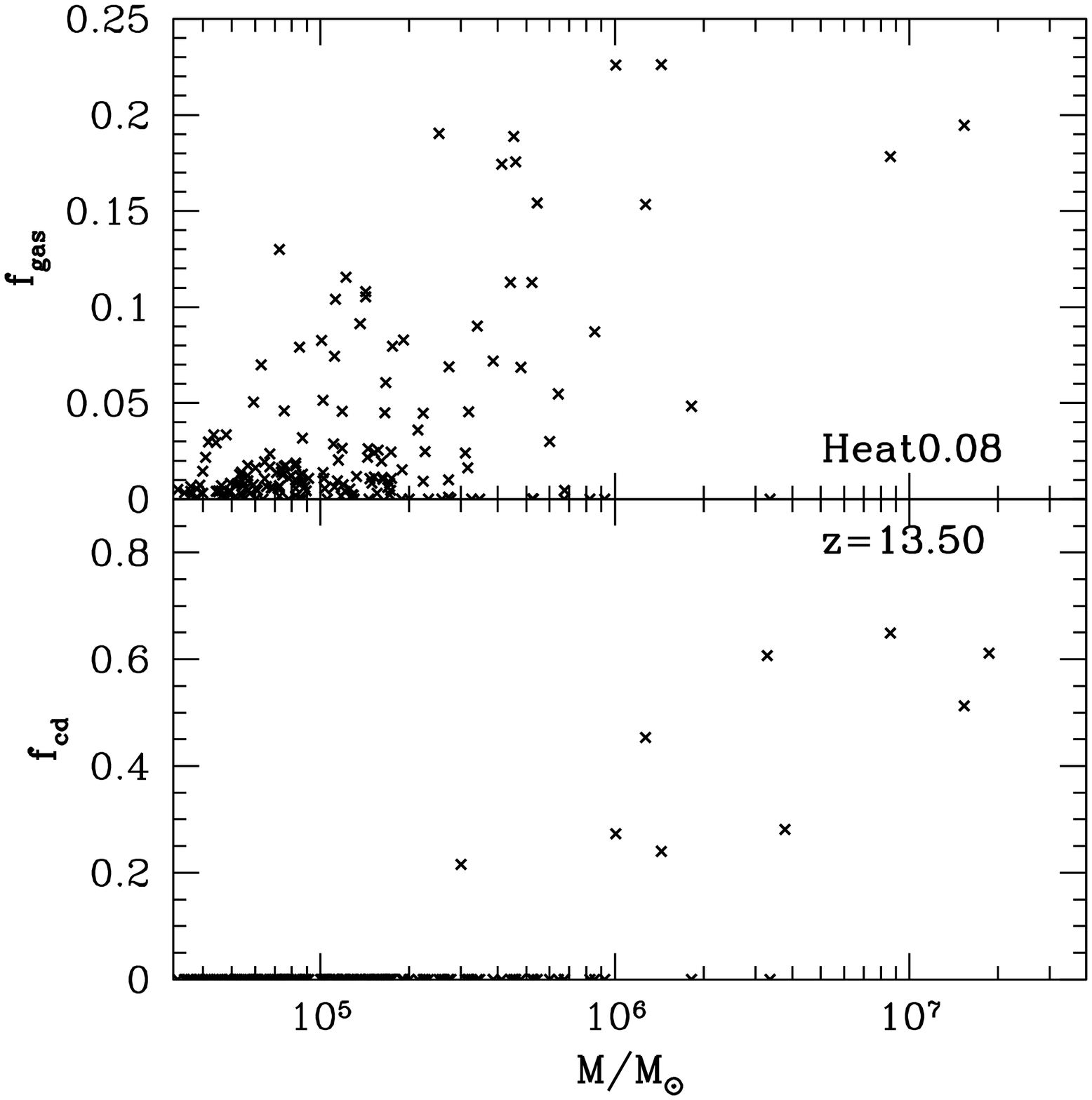}
\includegraphics[width=0.45\textwidth]{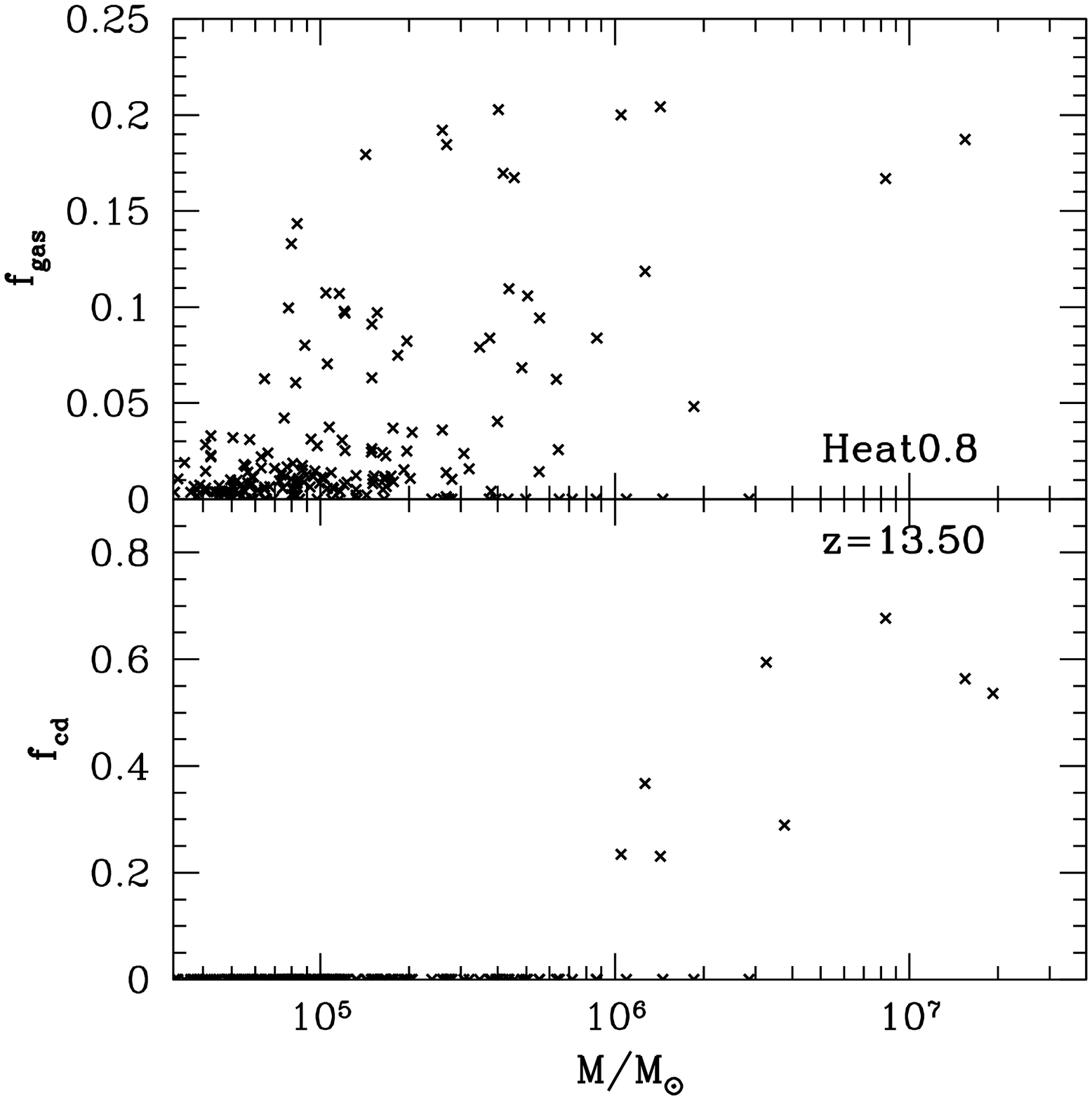}
}
\caption{
\label{fig:fcoll}
Total gas fractions ({\it top panels}) and cold, dense
gas fractions ({\it bottom panels}) as a function of total halo mass
at redshift $z=13.5$, the lowest redshift output of our simulations.
The four panels correspond to the \nothing\ ({\it top left}), \flash\
({\it top right}), \heatinglow\ ({\it bottom left}), and \heatinghigh\
({\it bottom right}) simulation runs. Note that all runs with a UVB have more halos with CD gas than the \nothing\ run.}
\vspace{+1.5\baselineskip}
\end{figure*}

In Fig. \ref{fig:fcoll}, we show the total gas fractions ({\it top panels}) and CD
gas fractions ({\it bottom panels}) from halos at $z=13.5$.  The positive feedback mentioned above is immediately evident from this figure, and shows explicitly the masses of the affected halos.  In particular, the \nothing\ run at $z=13.5$ has 3 {\it fewer} halos hosting CD gas than the \flash\ and \heatinglow\ runs, and 2 {\it fewer} than even the \heatinghigh\ run (see also Fig. \ref{fig:NM}).  These are some of the smallest and youngest of such halos at these redshifts.  We also verify that these halos formed in less biased and less overdense regions than the halos which didn't experience such positive feedback.

What is the physical cause of such delayed positive feedback?  Before we attempt to answer this question, let us briefly review the sequence of feedback-related events arising from a transient UVB:
\begin{enumerate}
\item When the UVB turns on, gas gets ionized and heated to temperatures $\gsim10^4$ K.
\item The temperature increase sets-off an outward moving pressure shock in the cores of halos, where density profiles have already steepened as gas approached hydrostatic equilibrium.
  This pressure shock smooths out the gas distribution and leads to a decrease in gas density in the cores of halos.
\item Once the UVB is turned off, the gas rapidly cools to $T\sim10^3$ K through a combination of atomic, molecular hydrogen and Compton cooling.  This temperature approximately corresponds to the gas temperature at the virial radius of such a proto-galactic, molecularly-cooled halo, thus effectively neutralizing the impact of temperature change on feedback.
\item A large amount of molecular hydrogen is produced, $x_{\rm H_2} \sim$ few $\times 10^{-3}$, irrespective of the gas density and temperature.
\item The pressure-shock begins to dissipate and gas with a newly enhanced H$_2$ abundance starts falling back onto the partially evacuated halo.
\end{enumerate}

It is the balance between the density depletion in step (ii) (which is dependent on the UVB intensity) and the enhanced cooling capability of the gas from step (iv) (which is independent of the UVB intensity), which determines the transient feedback outcomes discussed in \S~\ref{sec:trans}.  But what happens to gas which is not in the cores of already-formed halos with a steepened density profile?  In the extreme case of homogeneous gas, the UVB-induced temperature increase in (i) {\it does not translate into a pressure shock}.  Thus, there is no density depletion from step (ii).  In the more realistic case of a clumpy gas distribution, pressure gradients will still form and smooth out the gas distribution, but the resulting depletion will be less than in the cores of halos in hydrostatic equilibrium.  However, such lower-density gas outside of halos will still be able to cool to temperatures of $\sim10^3$ K since the Compton cooling time is independent of density, and at these high redshifts is over a factor of two shorter than the recombination time at mean density [in fact, these lower density regions take longer to recombine and hence can Compton-cool for a longer period reaching lower temperatures \citep{OH02}; see also equations (2) and (3) in MBH06].  Likewise, step (iv) mentioned above also applies to such lower density gas, with the transient background resulting in an enhanced molecular hydrogen abundance. Thus, we expect gas which has not yet formed a self-similar (e.g. \citealt{NFW96}) halo profile to experience weaker negative feedback [i.e. step (ii)], and the same amount of positive feedback [i.e. step (iv)], when exposed to identical transient UVBs.

This conclusion is perhaps a bit counterintuitive, since many cosmological radiative feedback studies find that feedback is a strong function of halo mass, with more massive, earlier forming halos being {\it less} susceptible to negative feedback \citep{Efstathiou92, BL99, Gnedin00filter, SIR04, KI00, Dijkstra04, MD08}.  However, there are two important distinctions here: we are discussing feedback resulting from a {\it transient} UVB; and we are comparing halos which have not yet formed and set up a self-similar gas profile at $\zuvboff$ to those which have already formed, rather than comparing {\it within} each of these two groups.  

\begin{figure*}
\vspace{+0\baselineskip}
{
\includegraphics[width=0.45\textwidth]{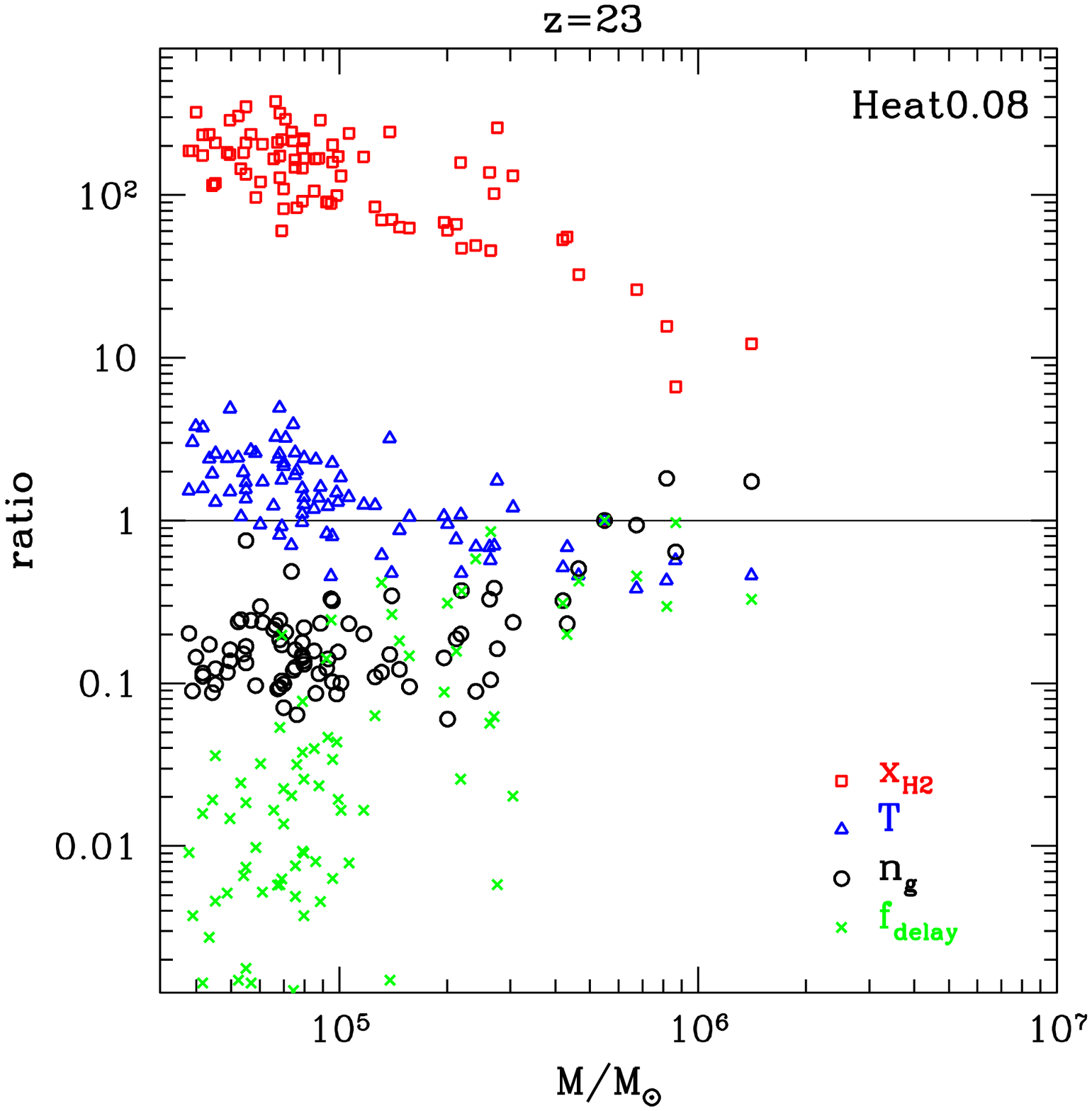}
\includegraphics[width=0.45\textwidth]{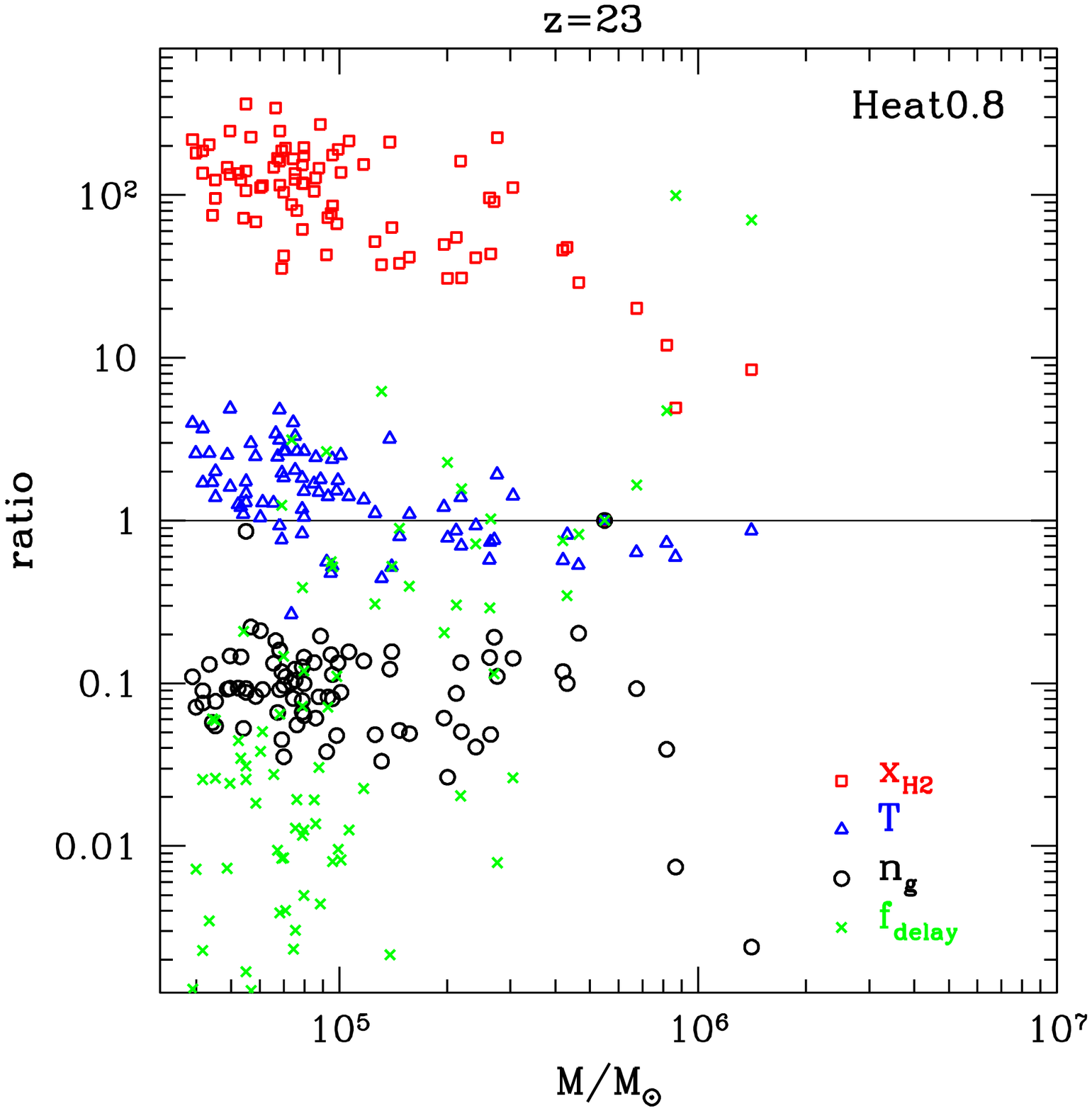}
}
\caption{
\label{fig:ratios}
Ratio of the ${\rm H_2}$ fractions ({\it red squares}), the temperatures ({\it blue
triangles}), and the average gas densities ({\it black circles}) within the central 15pc (proper) of halos.  Ratios of these quantities in the \heatinglow\ and \heatinghigh\ runs over the fiducial \nothing\ run are shown in the left and right panels, respectively, at $z=23$.  The delay factors from eq. (\ref{eq:delay}) are also plotted as green crosses.}
\vspace{+1.5\baselineskip}
\end{figure*}

We also see some evidence of these trends in Fig. \ref{fig:ratios}, where we plot the ratio (in runs with a UVB over the fiducial \nothing\ run) of the ${\rm H_2}$ fractions ({\it red squares}), the temperatures ({\it blue triangles}), and the average gas densities ({\it black circles}) within the central 15pc (proper) of halos.  
These quantities are shown shortly after $\zuvboff$, at $z=23$.  As already noted in \S \ref{sec:trans} and in (iv) above, the ${\rm H_2}$ fraction tends to increase by the same factor, regardless of the nature of the heating. As a result, the sign of the overall feedback (negative or positive) is determined primarily by whether or not the depression of the gas density overcomes the positive effects of this H$_2$ enhancement. The density depletion does depend strongly on the type and amount of heating.
  This depletion, shown in black circles, is of course stronger for the \heatinghigh\ than the \heatinglow\ run.  Thus halos in the former initially experience strong negative feedback and in the latter negligible or weak positive feedback, as we've seen in \S \ref{sec:trans} and as noted by the green crosses in the figure.  Note however that the difference in the amount of gas depletion between these two heating runs decreases with decreasing halo mass as one begins being affected by Jeans/filtering scale effects\footnote{Note that accurately modeling Jeans smoothing requires very high resolution in both dark matter and gas \citep{NBM09}. These small scales are shown in Fig. \ref{fig:ratios} merely to show qualitative trends.}. Following the above argument, this trend should continue and density depletion should decrease in regions which are less evolved (i.e. the black points should approach unity as the halo mass is decreased below the values shown in the figure).  It is these regions which become the progenitors of our halos experiencing eventual positive feedback.
  Unfortunately, at $z=23$, these regions are too unevolved to be identified by our halo finding algorithm and thus do not appear in Fig. \ref{fig:ratios}.

This eventual positive feedback should be quite sensitive to the LWB.  Indeed, looking at the bottom panel of Fig. \ref{fig:LW_delta}, we see that this feedback effect is wiped-out if $\Jlwb\gsim10^{-3}$.  It is likely that such modest values of $\Jlwb$ are reached well before the majority of the Universe is ionized \citep{HAR00}, though the buildup of the LWB can be slowed by an increasing optical depth to LW photons from the enhanced H$_2$ abundances in relic HII regions \citep{JGB07}, or in shells surrounding HII regions (\citealt{RGS01, KM05}; though see \citealt{KYSU04}).  Thus studying the true physical importance of this positive feedback mechanism would have to be done with self-consistent cosmological simulations of reionization.

\begin{figure*}
\vspace{+0\baselineskip}
{
\includegraphics[width=0.245\textwidth]{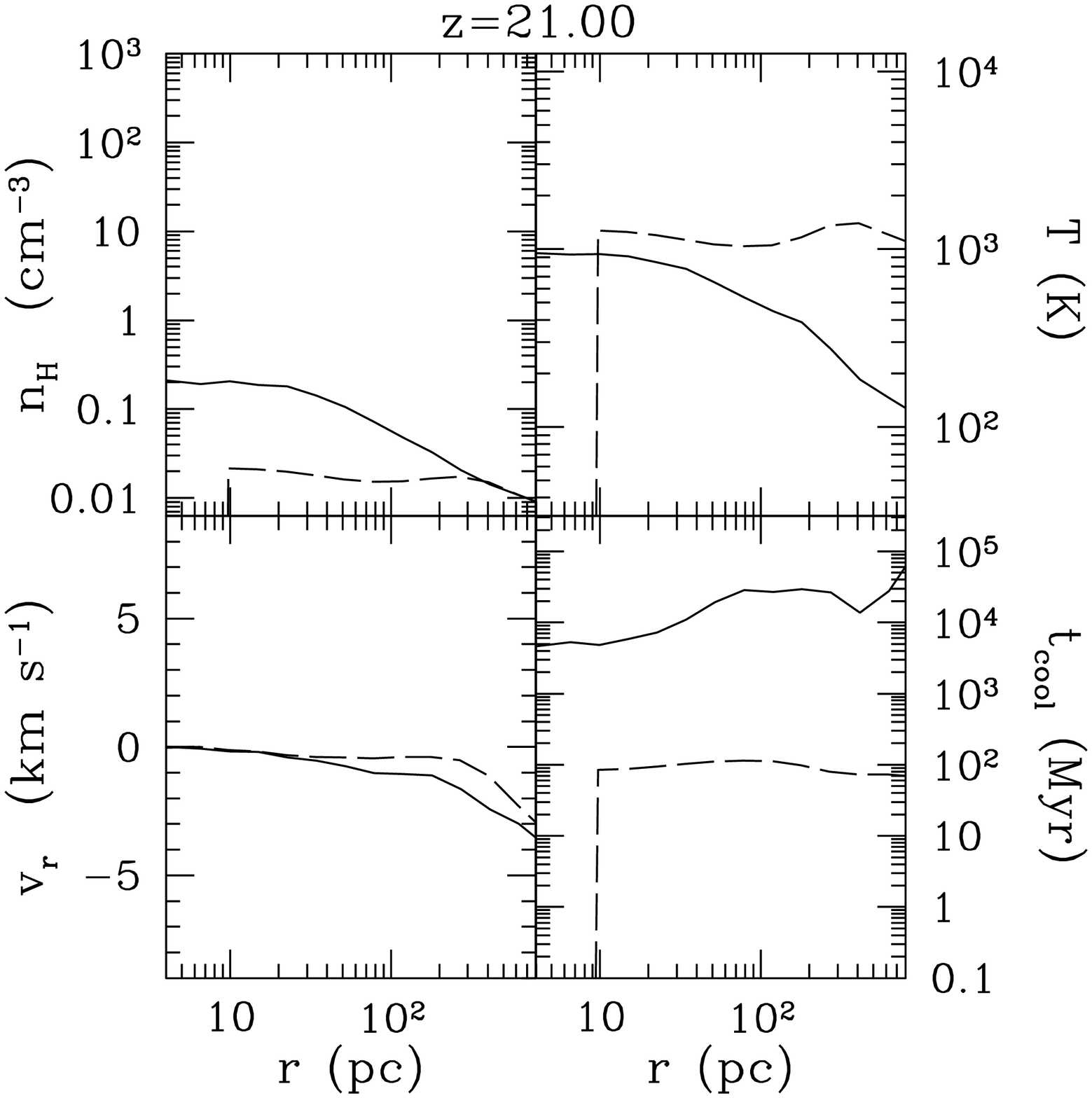}
\includegraphics[width=0.245\textwidth]{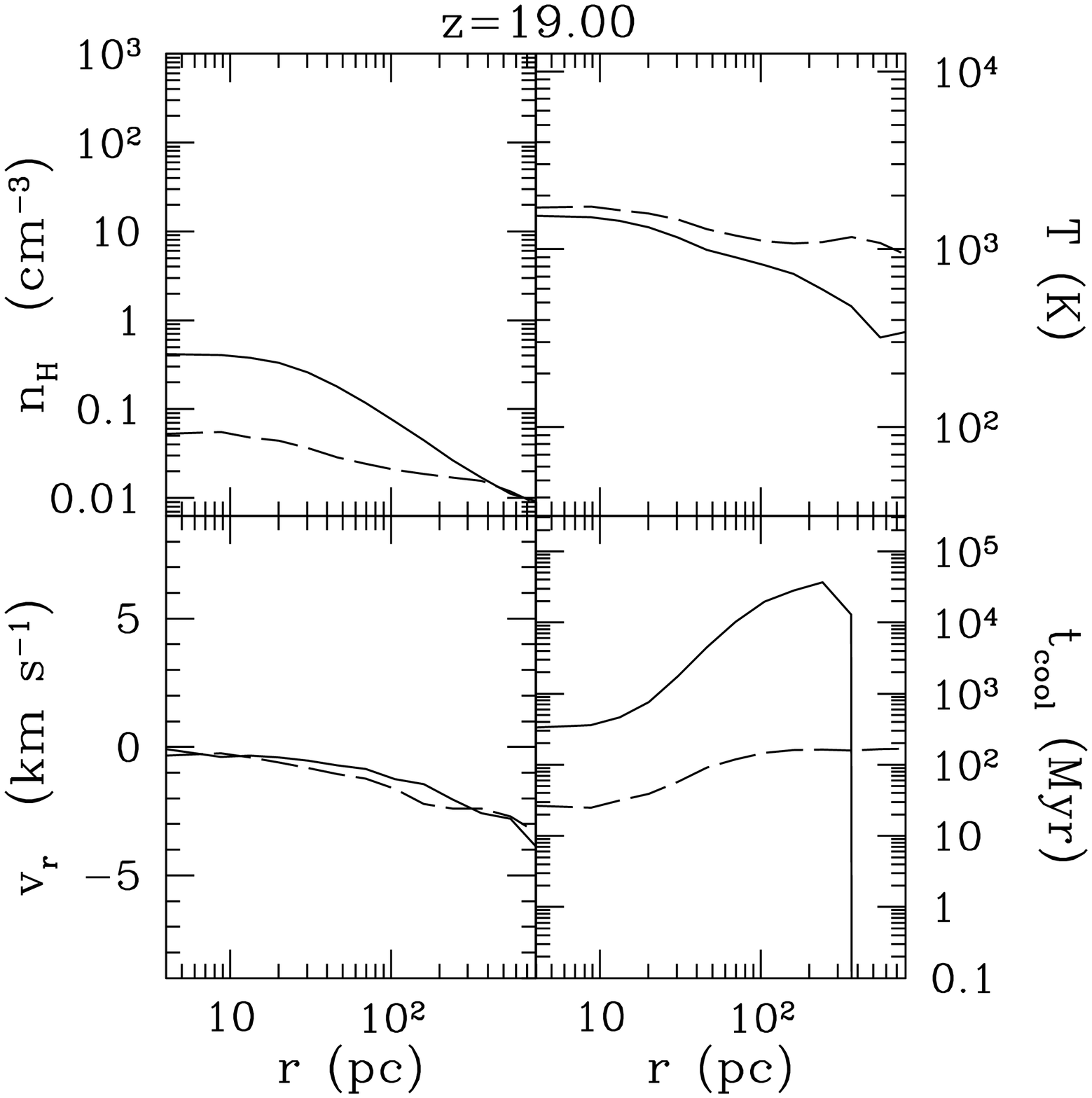}
\includegraphics[width=0.245\textwidth]{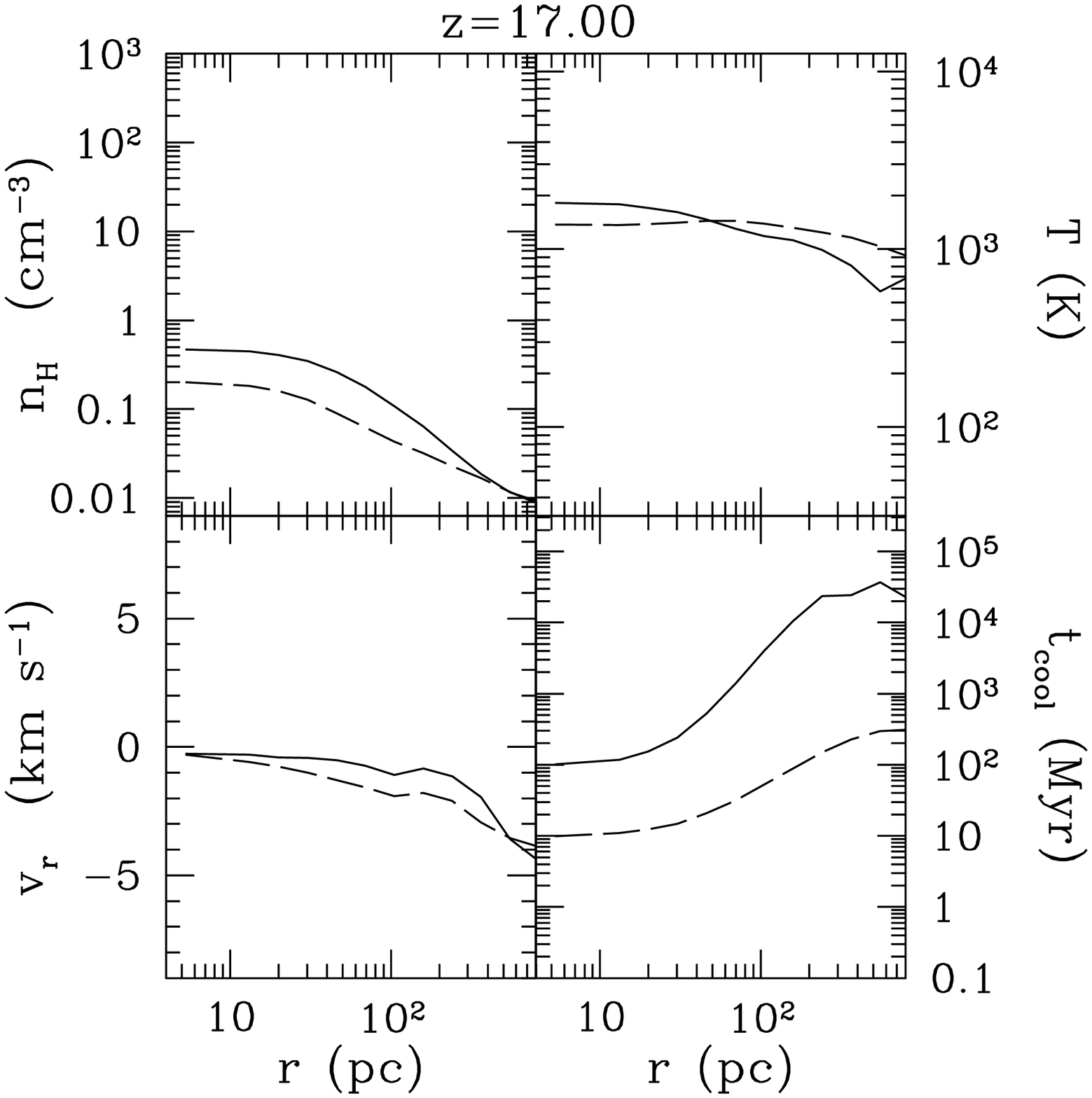}
\includegraphics[width=0.245\textwidth]{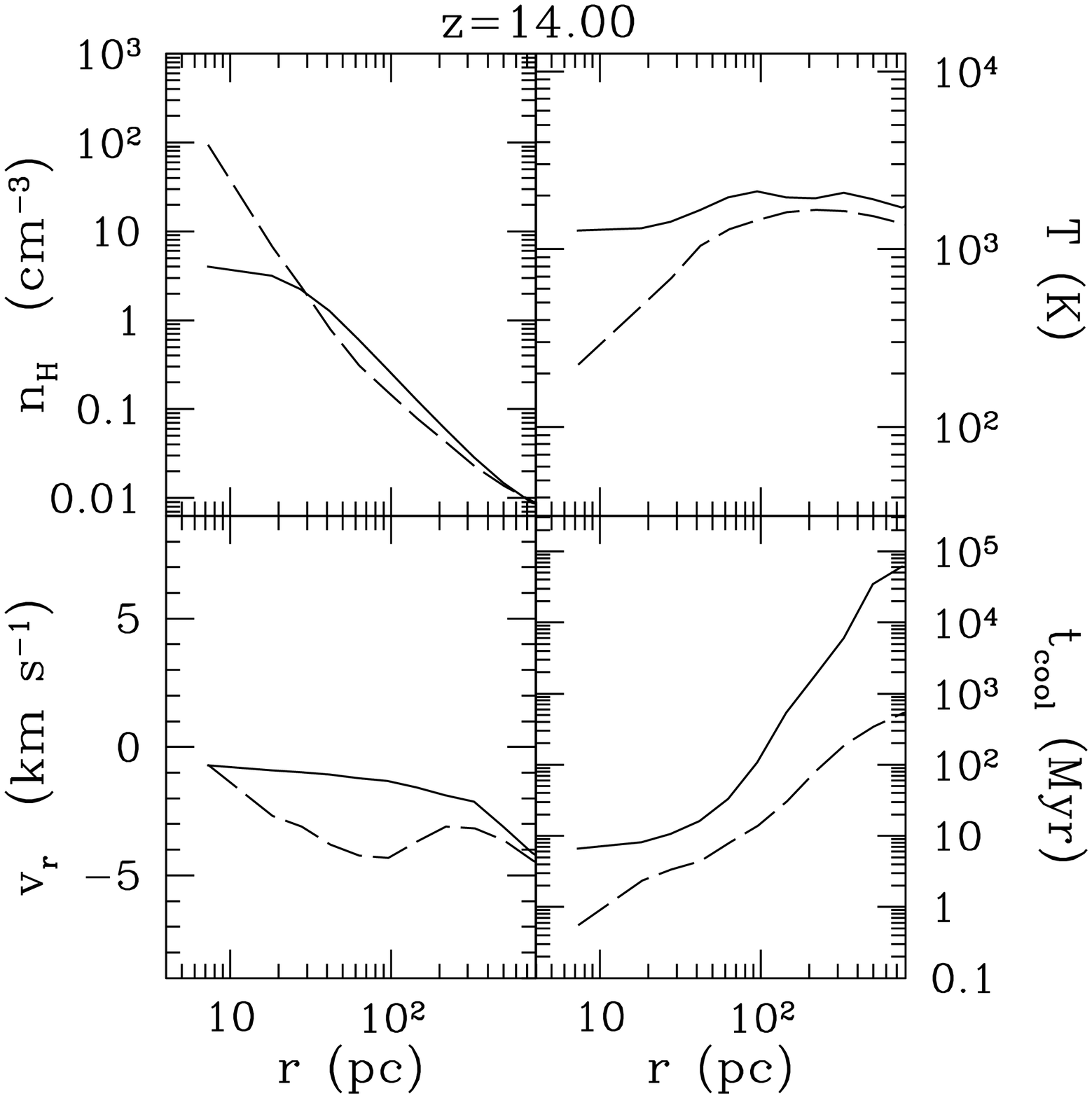}
\vskip0.0pt
}
{
\includegraphics[width=0.245\textwidth]{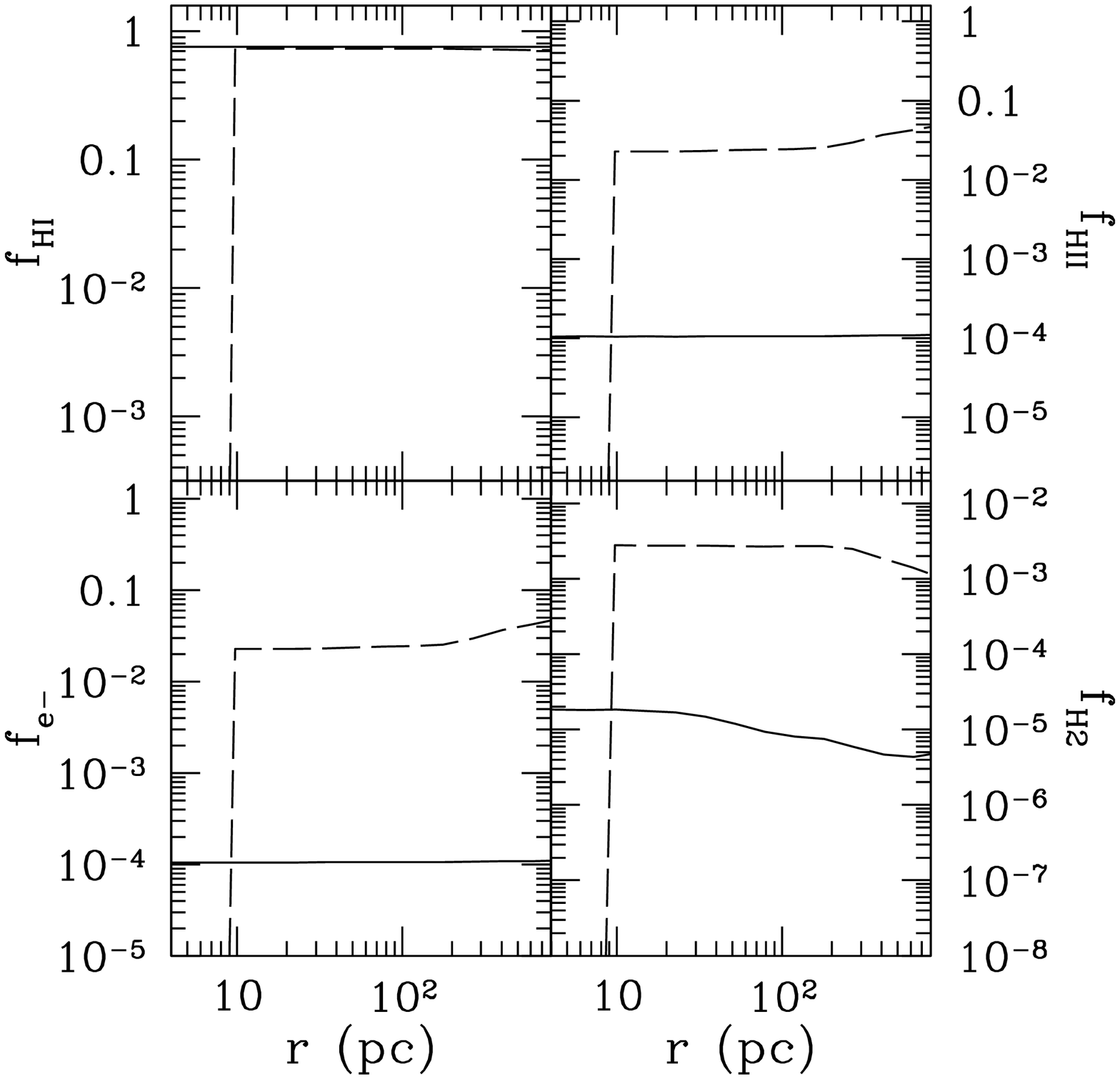}
\includegraphics[width=0.245\textwidth]{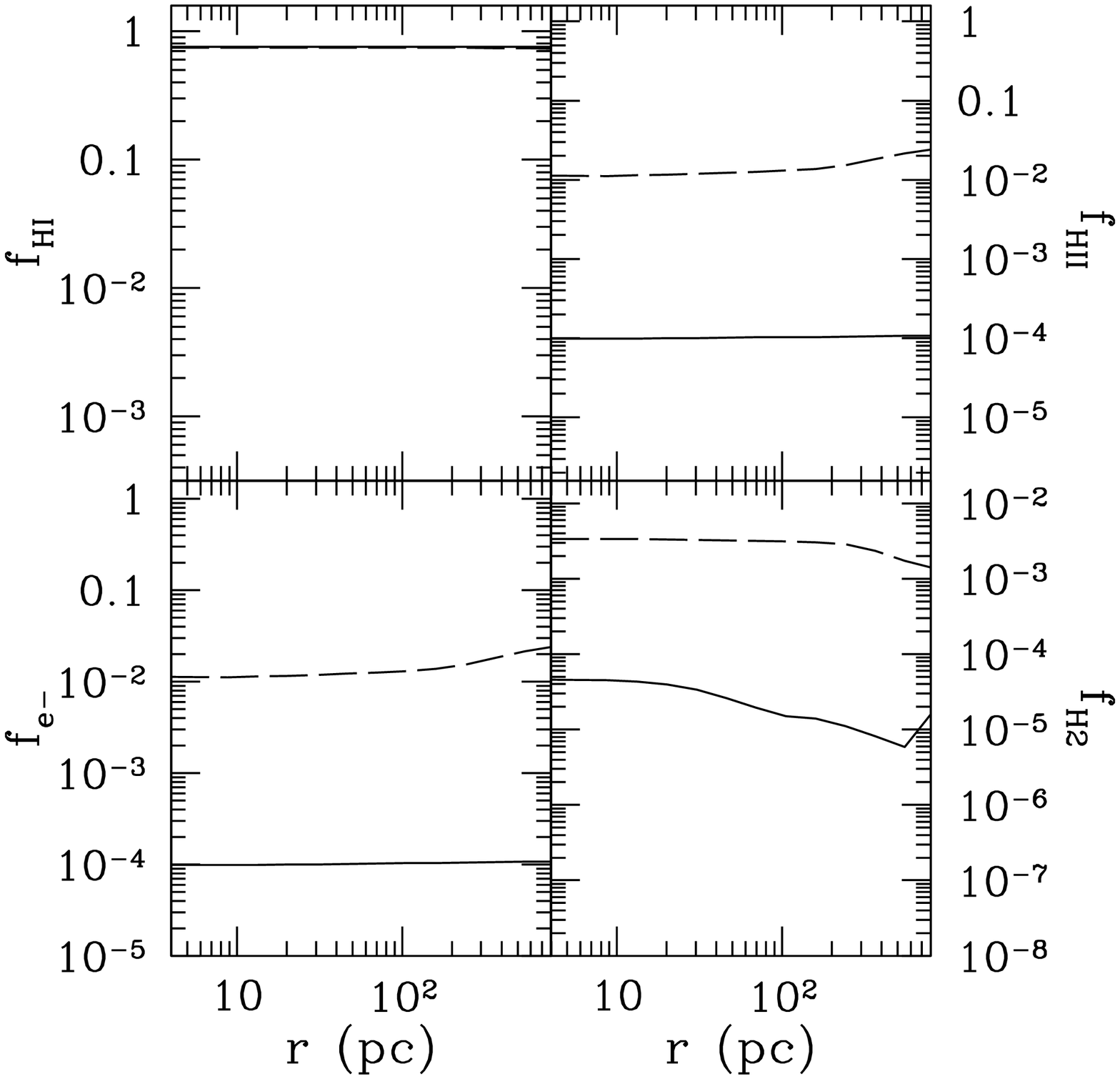}
\includegraphics[width=0.245\textwidth]{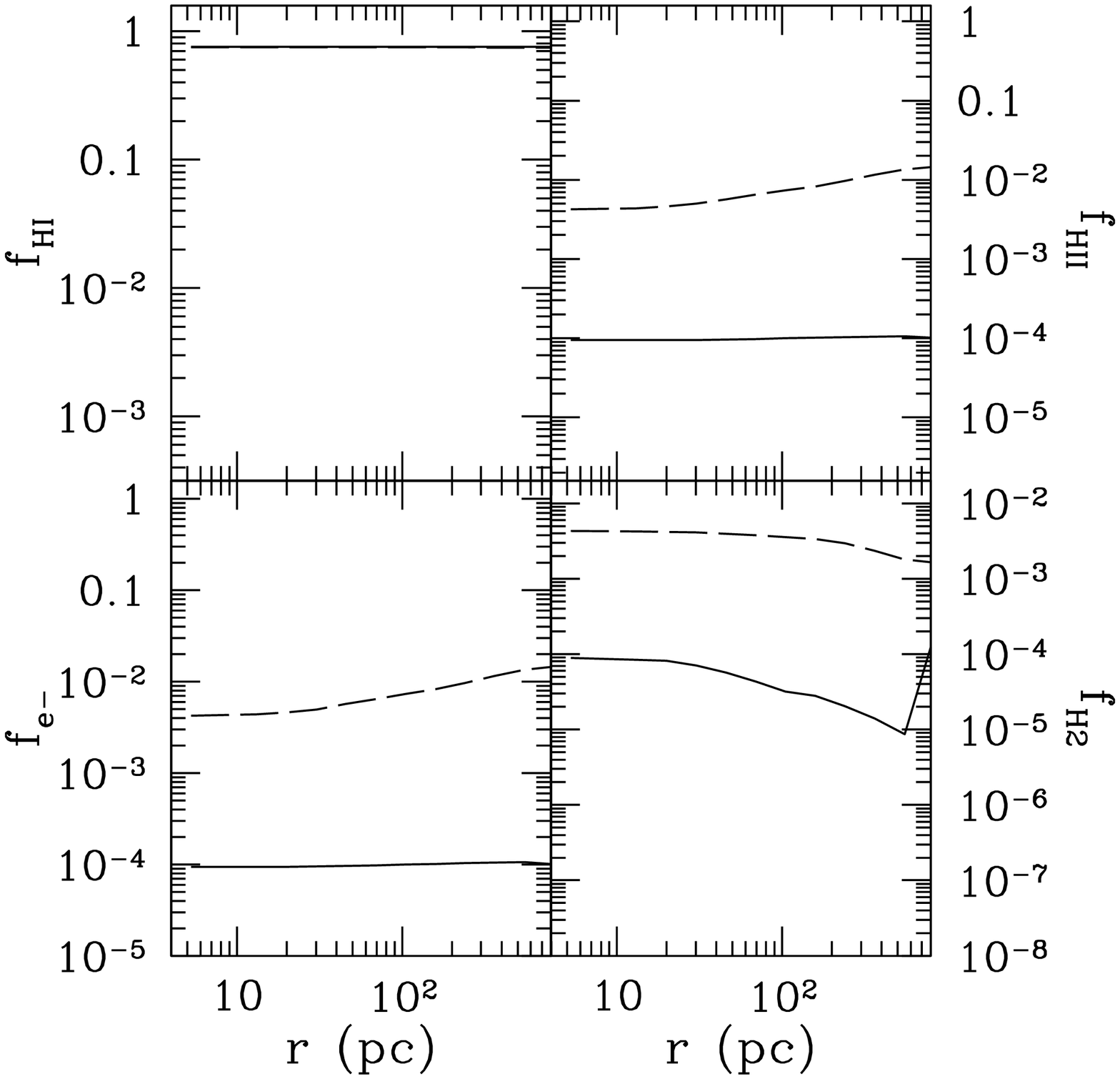}
\includegraphics[width=0.245\textwidth]{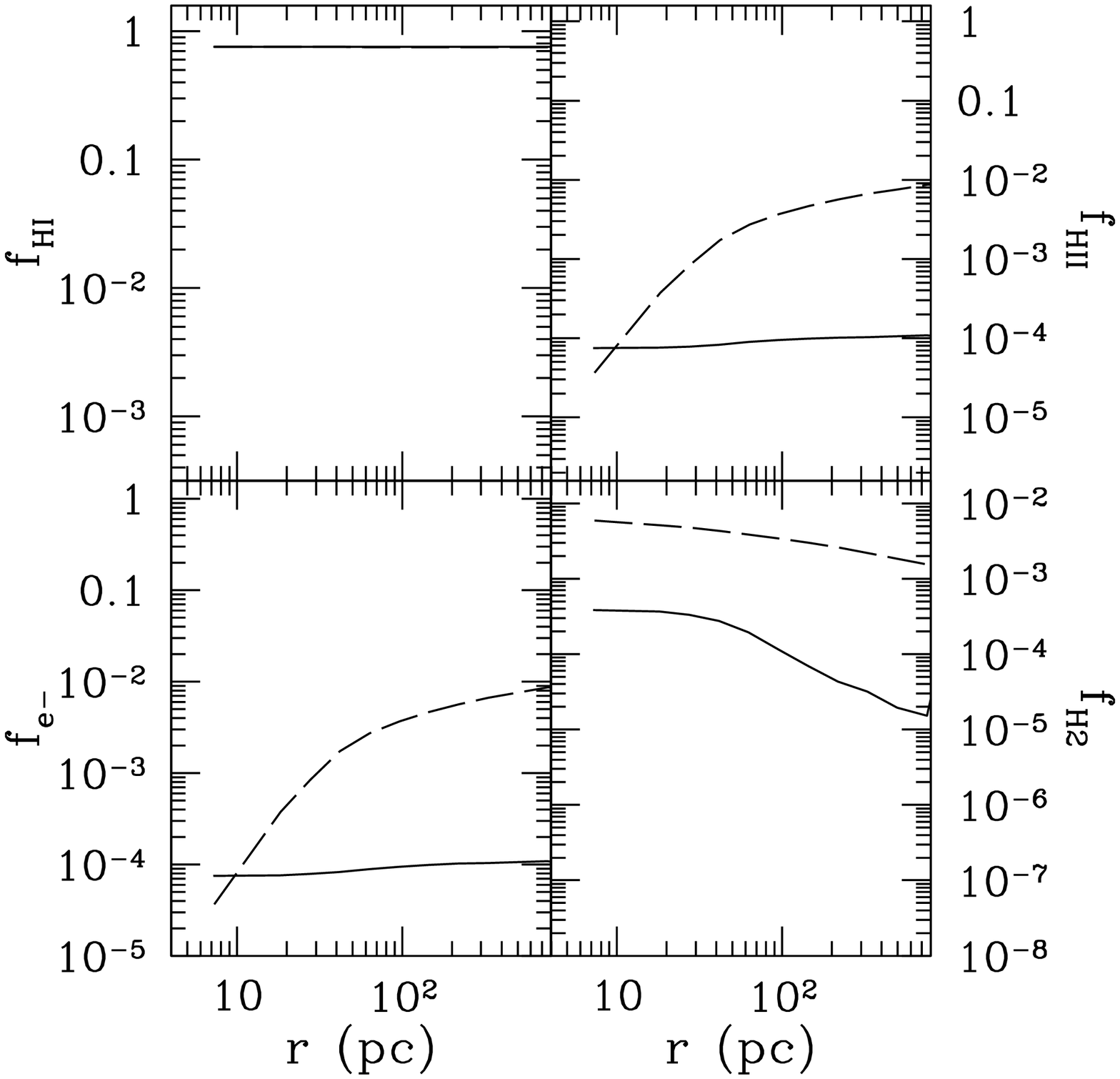}
}
 \caption{
The same quantities as in Fig. \ref{fig:early_profiles}, but for a halo which experiences positive feedback in the \heatinghigh\ run. Solid and dashed curves correspond to the \nothing\ and \heatinghigh\ runs, respectively.
\label{fig:NoUVB_vs_Heat}
}
\vspace{-1\baselineskip}
\end{figure*}

Now, let us look at specific examples by studying the evolution of one such halo experiencing eventual positive feedback.  In Figure \ref{fig:NoUVB_vs_Heat}, we plot the same quantities as in Fig. \ref{fig:early_profiles}, but for a halo which experiences positive feedback in the \heatinghigh\ run.  Note that, unlike the more evolved halos which are subjected to the transient feedback and have already formed at $\zuvboff$, the earliest snapshot we have of this halo is at $z=21$. Solid and dashed curves correspond to the \nothing\ and \heatinghigh\ runs, respectively.

It is important to note that this halo forms considerably after $\zuvboff$, and thus the ``density depletion" seen in the figure at $z\gsim 17$ is of a different nature than discussed in \S \ref{sec:trans} and Fig. \ref{fig:early_profiles}.  Namely, this gas wasn't evacuated from the halo center with a pressure shock.  Instead, these late forming halos start pulling in baryons when their virial temperature exceeds that of the gas temperature.  The heating and subsequent Compton cooling following the transient UVB sets the IGM gas temperature in these runs at $T\sim10^3$ K.  Thus baryonic infall onto newly formed halos can commence when $\Tvir \gsim 10^3$ K.  Since the IGM is cooler in the \nothing\ run, gas there gets a ``head-start'' in accreting onto DM halos.  However, as is evident in the figure, the enhanced H$_2$ cooling channel in the \heatinghigh\ run allows the gas to quickly catch-up and surpass its twin in the \nothing\ run.  Thus by redshift $z=14$, the halo has formed CD gas in the \heatinghigh\ run, but not in the \nothing\ run.

\begin{figure*}
\vspace{+0\baselineskip}
{
\includegraphics[width=0.245\textwidth]{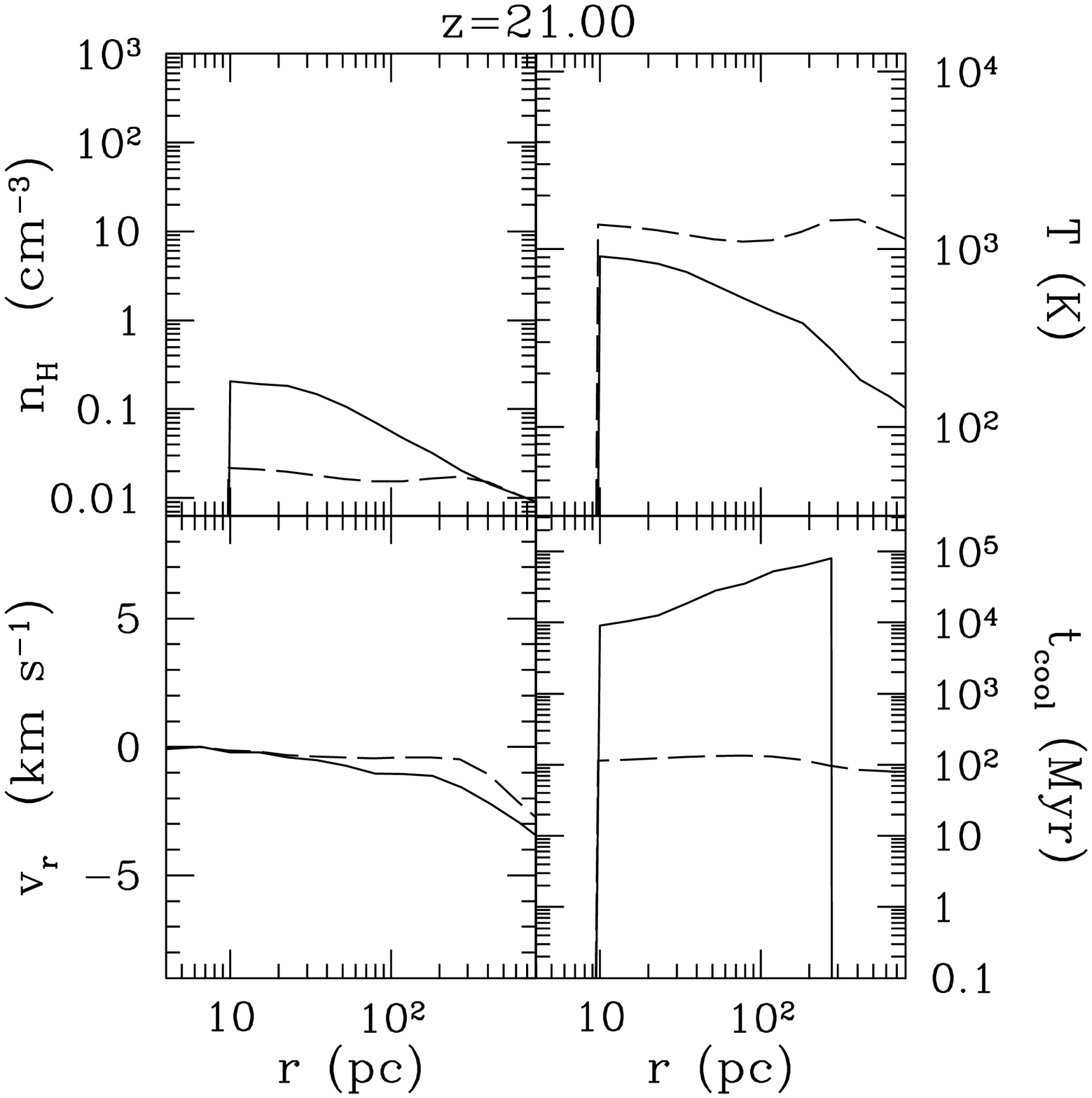}
\includegraphics[width=0.245\textwidth]{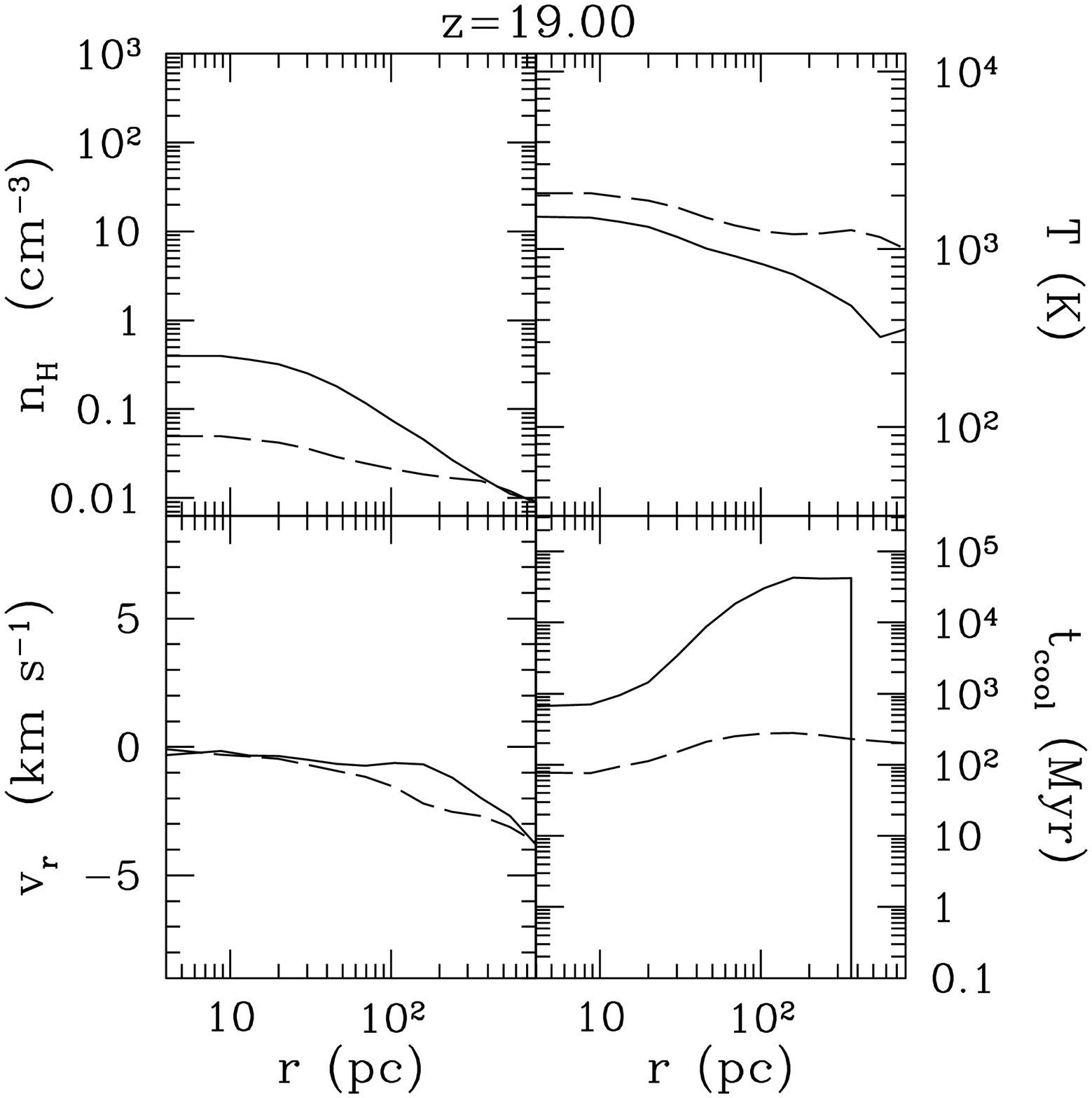}
\includegraphics[width=0.245\textwidth]{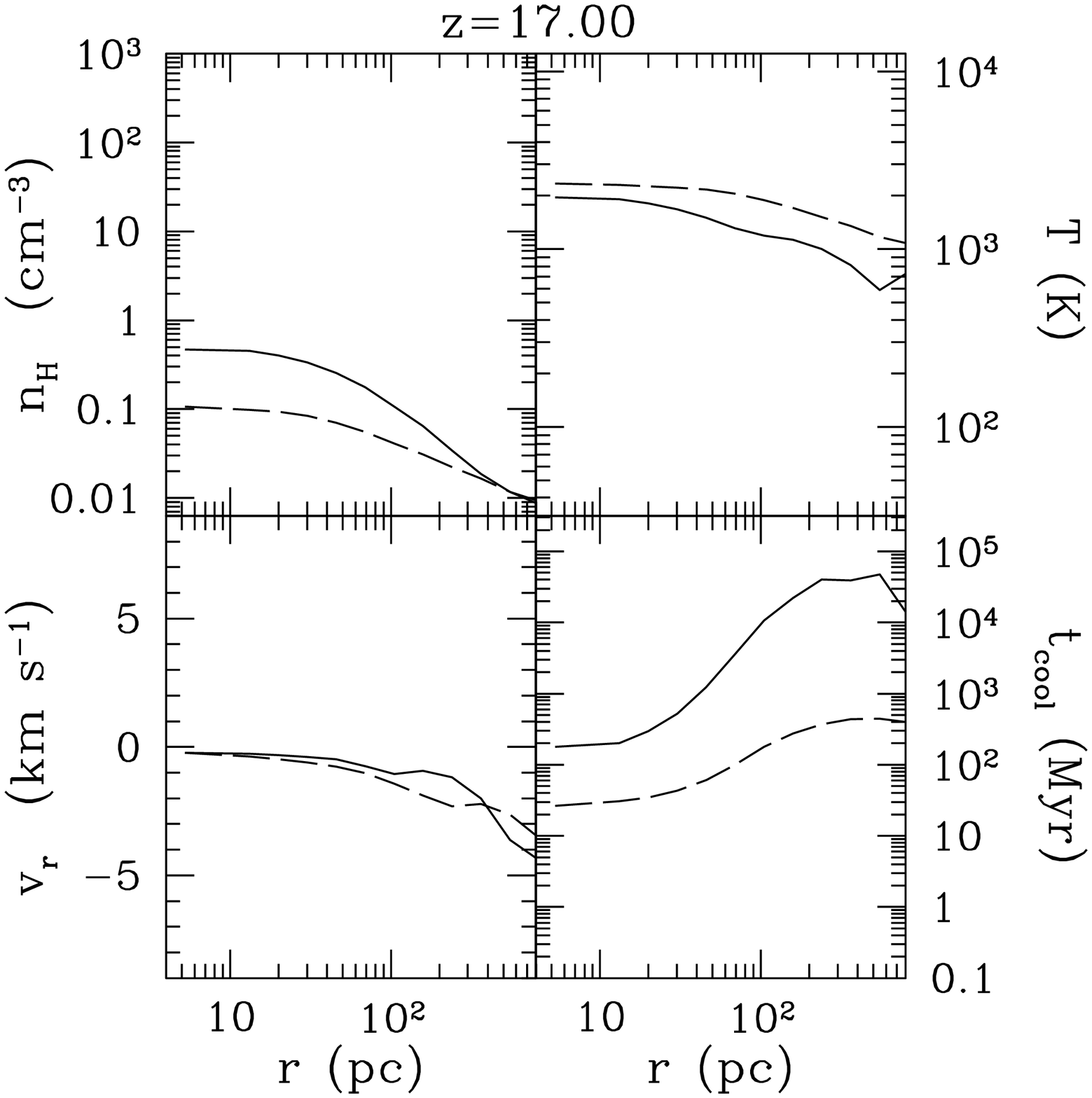}
\includegraphics[width=0.245\textwidth]{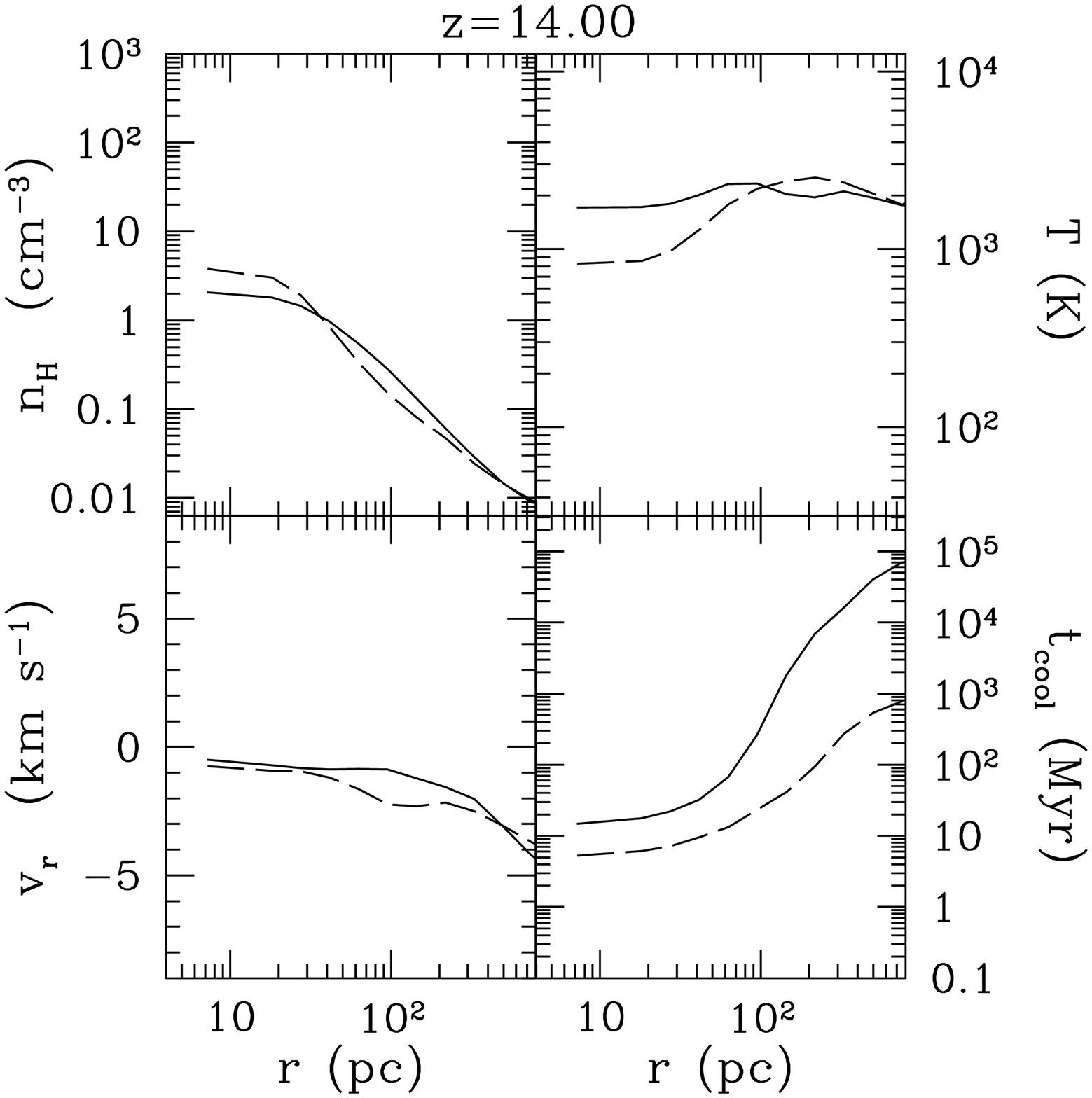}
\vskip0.0pt
}
{
\includegraphics[width=0.245\textwidth]{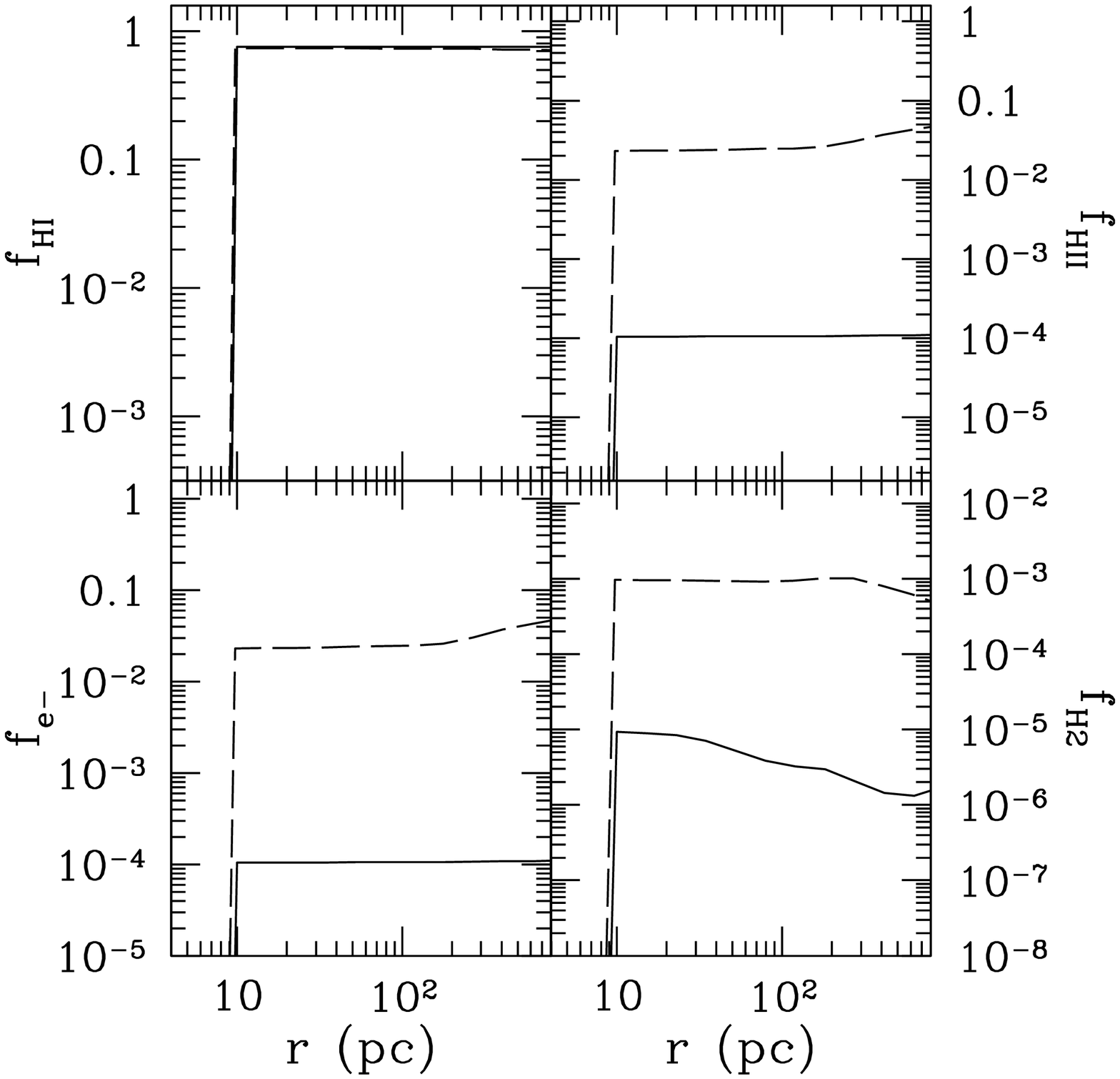}
\includegraphics[width=0.245\textwidth]{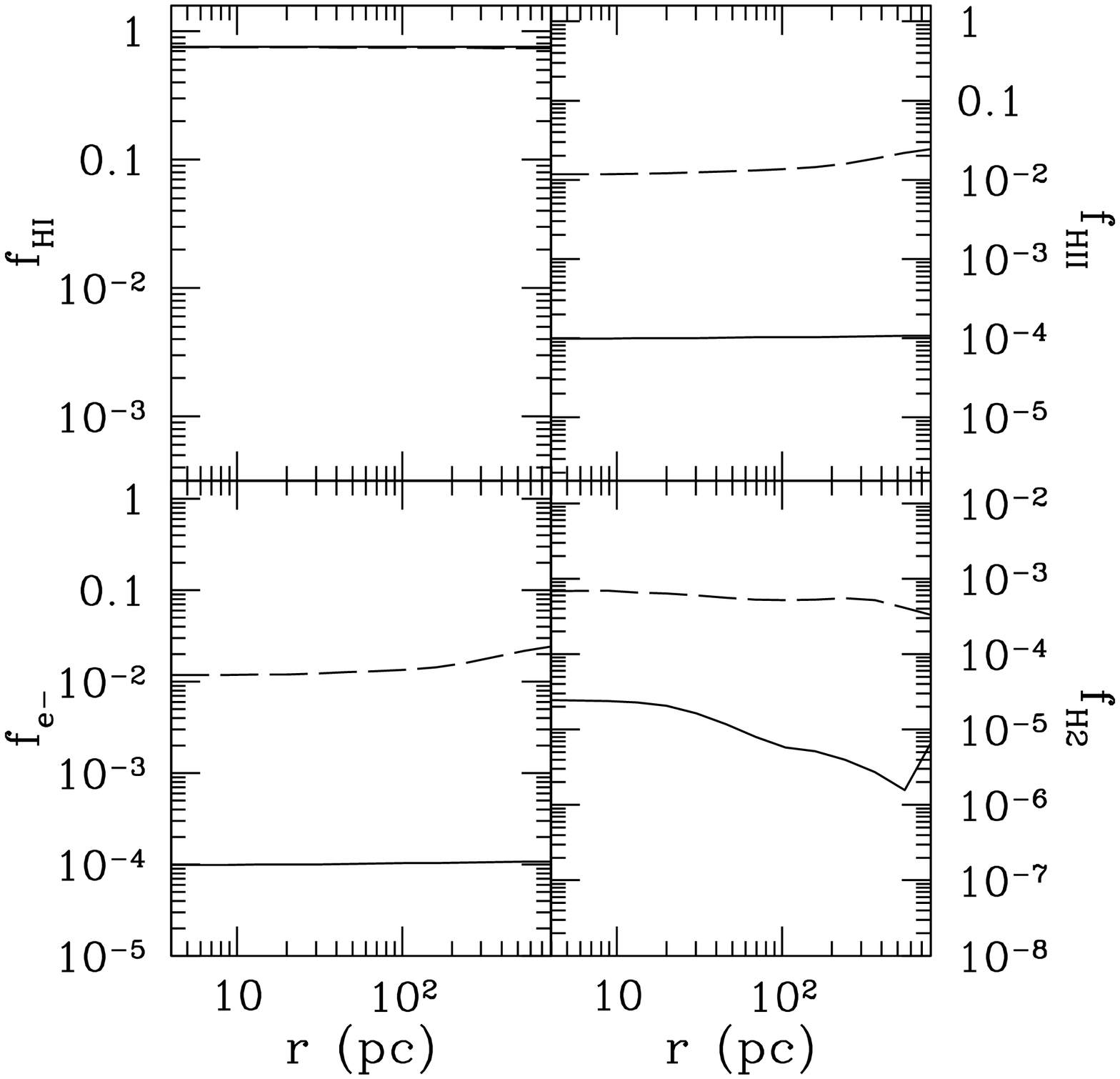}
\includegraphics[width=0.245\textwidth]{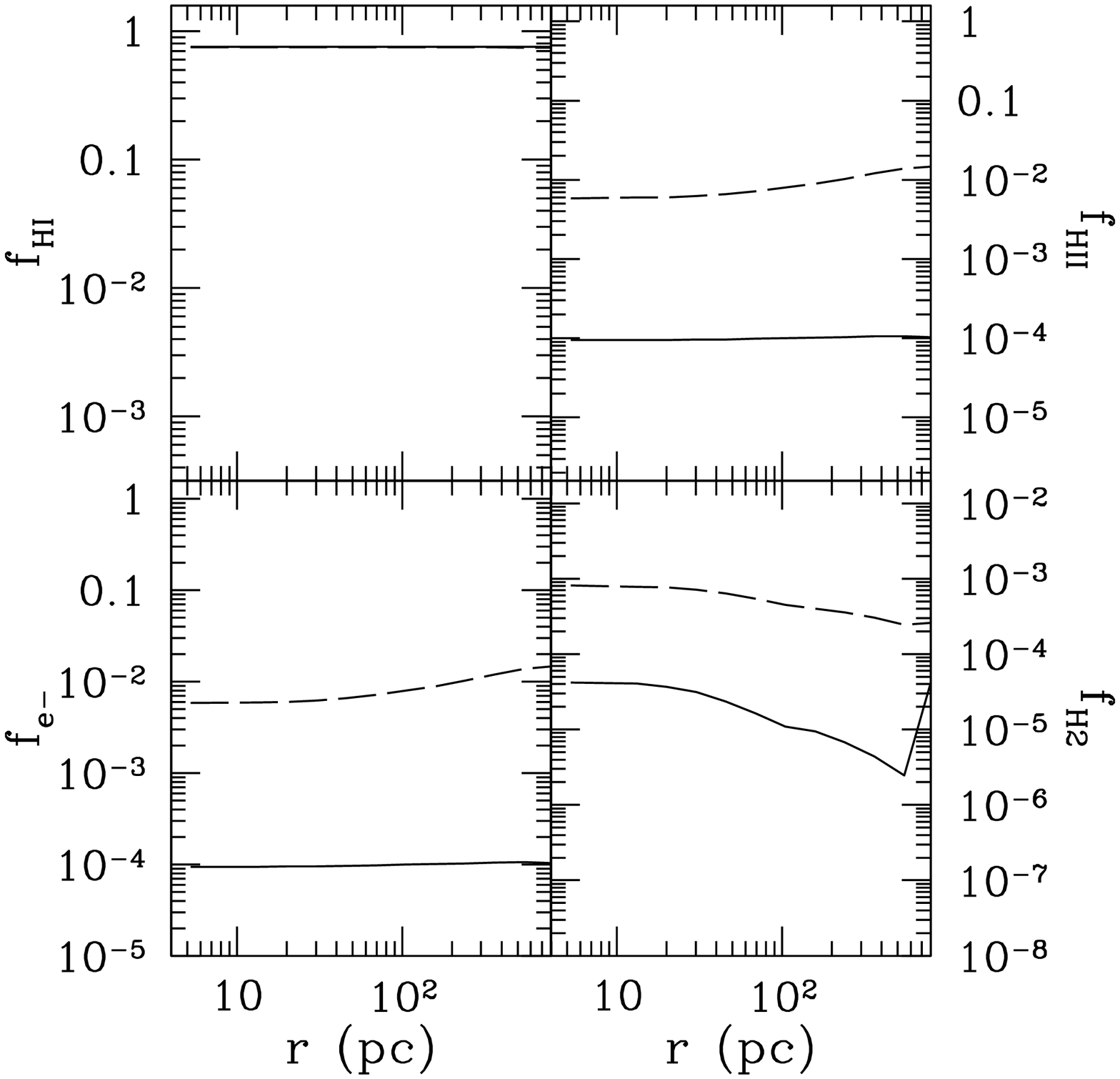}
\includegraphics[width=0.245\textwidth]{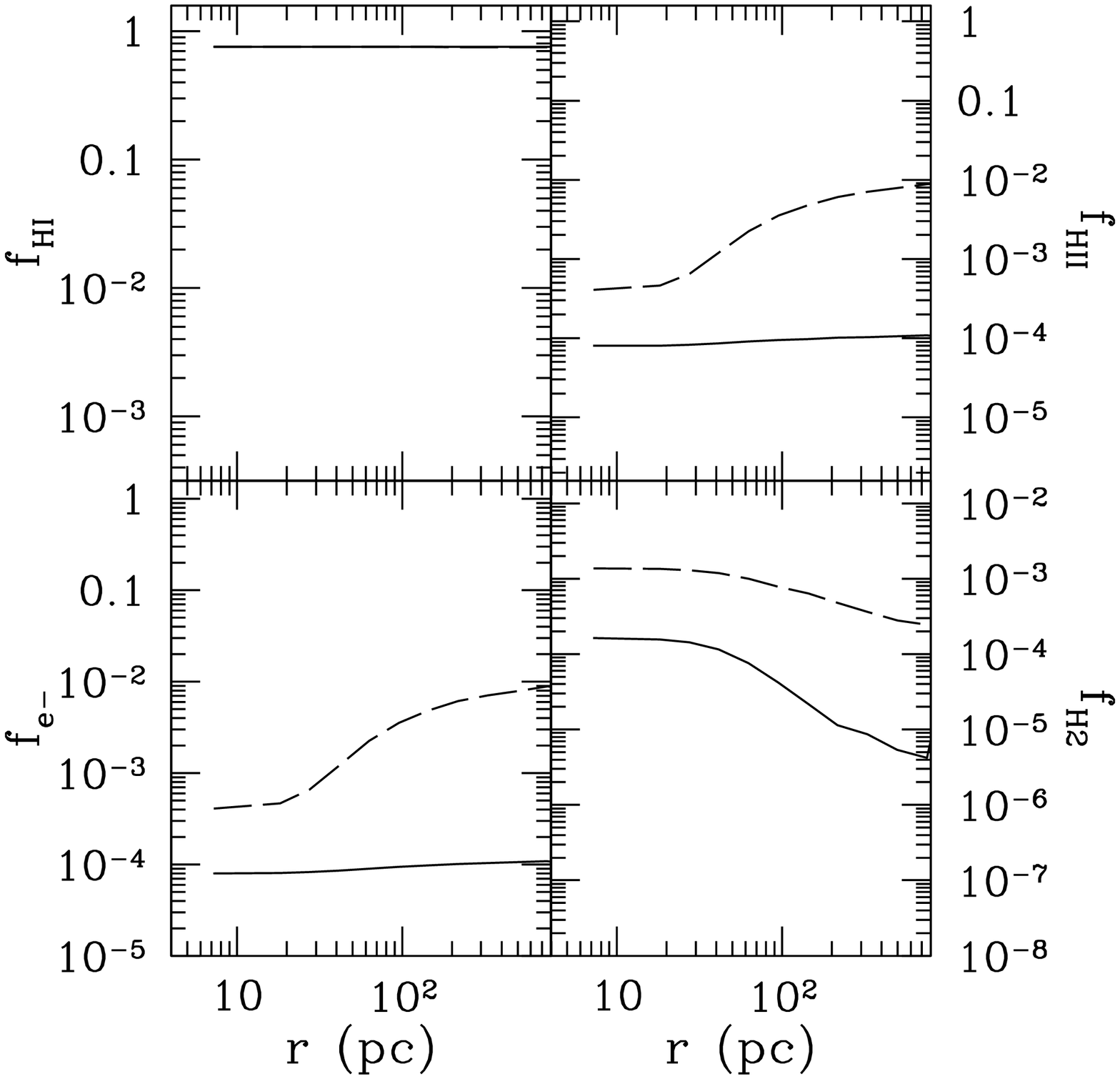}
}
 \caption{
Same as Fig. \ref{fig:NoUVB_vs_Heat}, but with solid and dashed curves corresponding to the \LWthreenothing\ and \LWthreeheatinghigh\ runs, respectively.
\label{fig:LW3NoUVB_vs_LW3Heat}
}
\vspace{-1\baselineskip}
\end{figure*}

How does adding a LWB impact this positive feedback?  From Fig. \ref{fig:LW_delta}, we have already surmised that a background intensity of $\Jlwb\gsim10^{-3}$, can offset these trends by decreasing the enhanced H$_2$ fraction in relic HII regions.  In Fig. \ref{fig:LW3NoUVB_vs_LW3Heat}, we plot profiles from the same halo in the \LWthreenothing\ ({\it solid curves}) and \LWthreeheatinghigh\ ({\it dashed curves}).  By comparing the solid curves from Fig. \ref{fig:NoUVB_vs_Heat} to the dashed curves from Fig. \ref{fig:LW3NoUVB_vs_LW3Heat}, one confirms  that a LWB with $\Jlwb\sim10^{-3}$ effectively neutralizes the positive feedback from the transient UVB.   Thus at $z=14$ the halo core has a similar gas profile and cooling time in the \nothing\ and \LWthreeheatinghigh\ runs.

By comparing the two profiles at $z=14$ in Fig. \ref{fig:LW3NoUVB_vs_LW3Heat}, we see that the transient UVB still stimulates positive feedback, but the effect is not nearly as dramatic.  Specifically, in the halo core at $z=14$ the density is enhanced and the cooling time is lowered (both by a factor of $\sim2$) in the \LWthreeheatinghigh\ run compared to the \LWthreenothing\ run.  This level of positive feedback is much more modest compared with the case with no LWB, where the equivalent density enhancement was a factor of $\sim50$ and the cooling time was lower by a factor of $\sim15$ (see Fig. \ref{fig:NoUVB_vs_Heat}).

\begin{figure}
\includegraphics[width=0.45\textwidth]{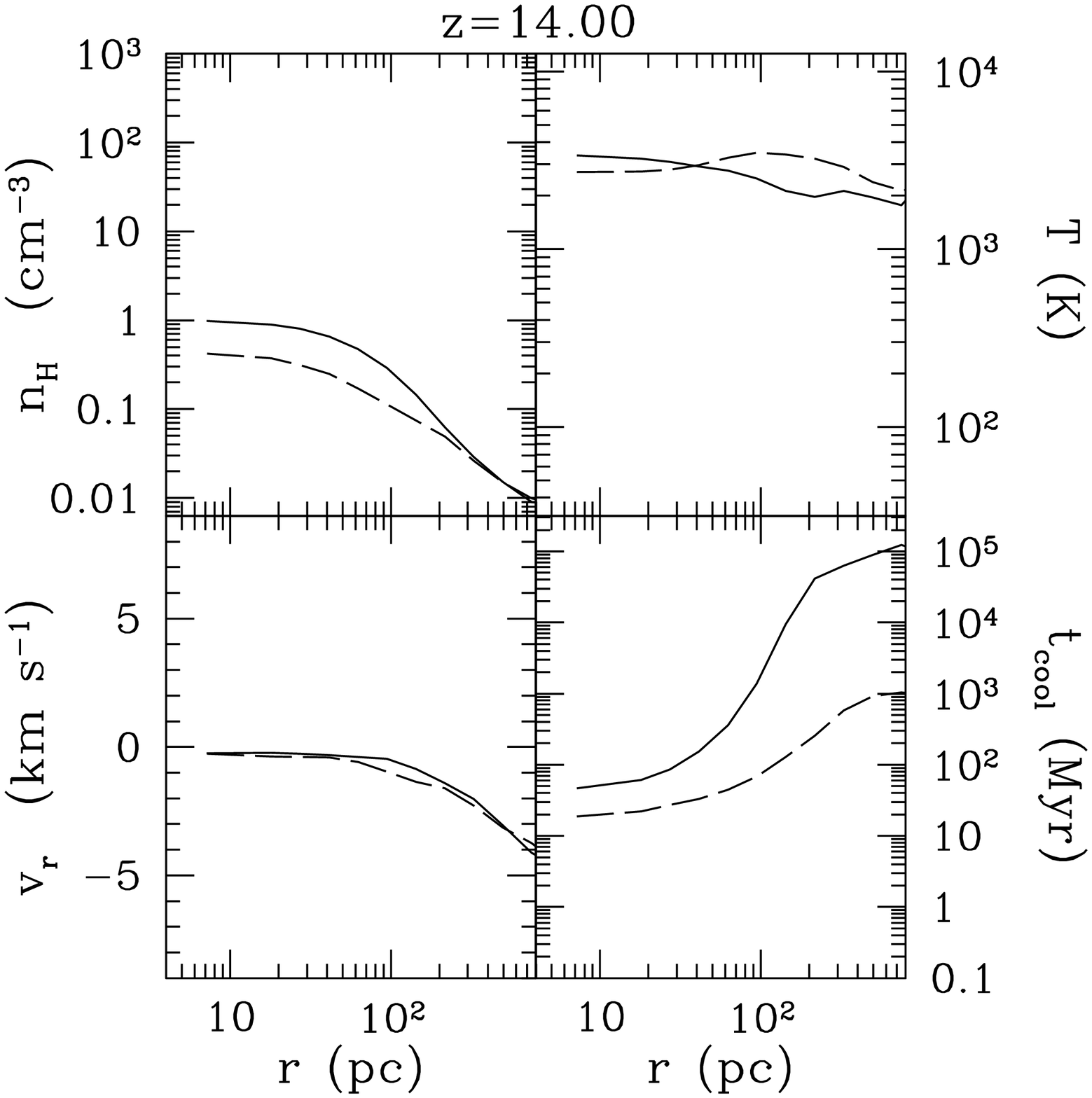}
\includegraphics[width=0.45\textwidth]{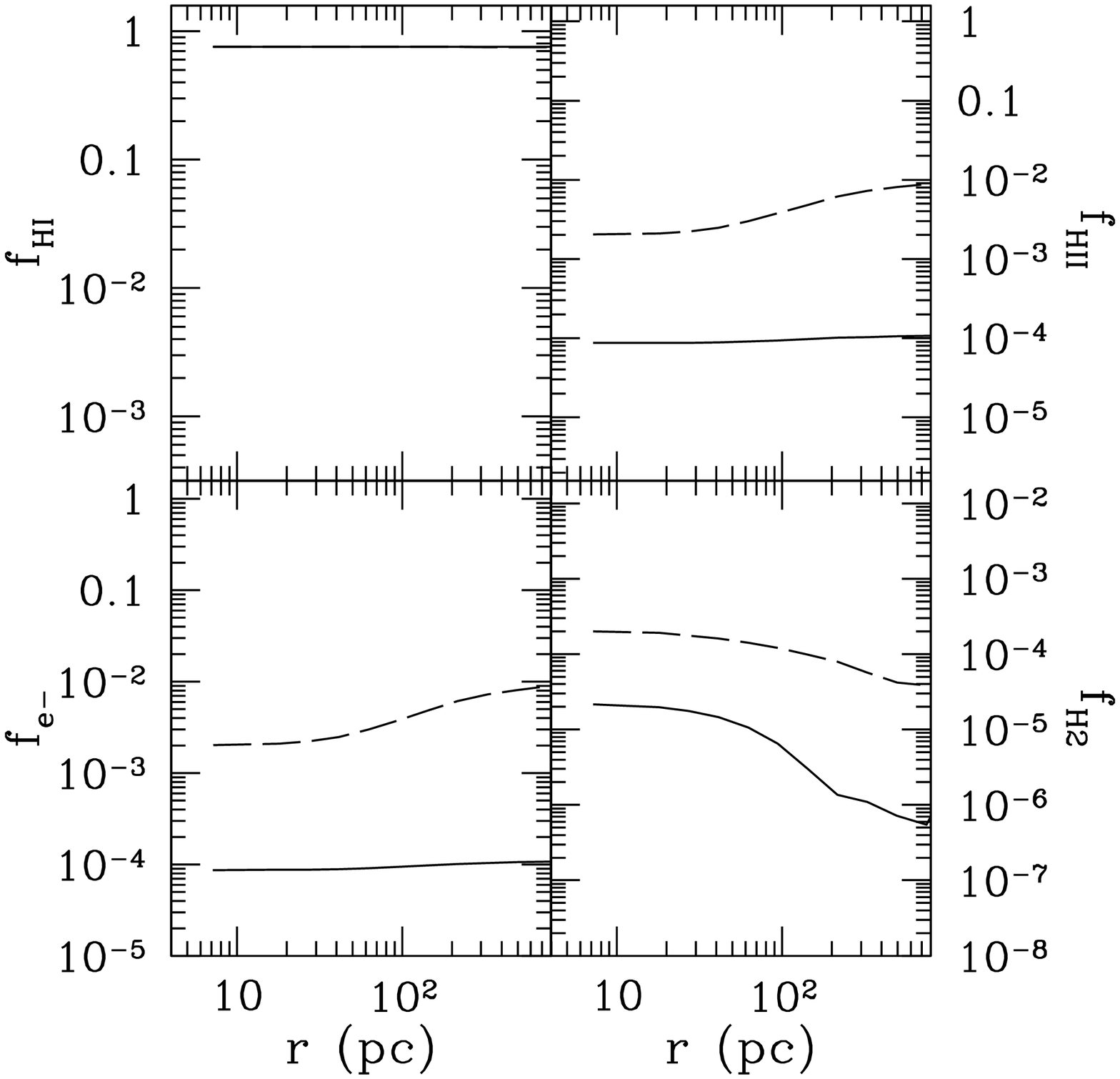}
 \caption{
Same as Fig. \ref{fig:NoUVB_vs_Heat}, at $z=14$, but with solid and dashed curves corresponding to the \LWtwonothing\ and \LWtwoheatinghigh\ runs, respectively.
\label{fig:LW2NoUVB_vs_LW2Heat}
}
\vspace{-1\baselineskip}
\end{figure}

\begin{figure}
\includegraphics[width=0.45\textwidth]{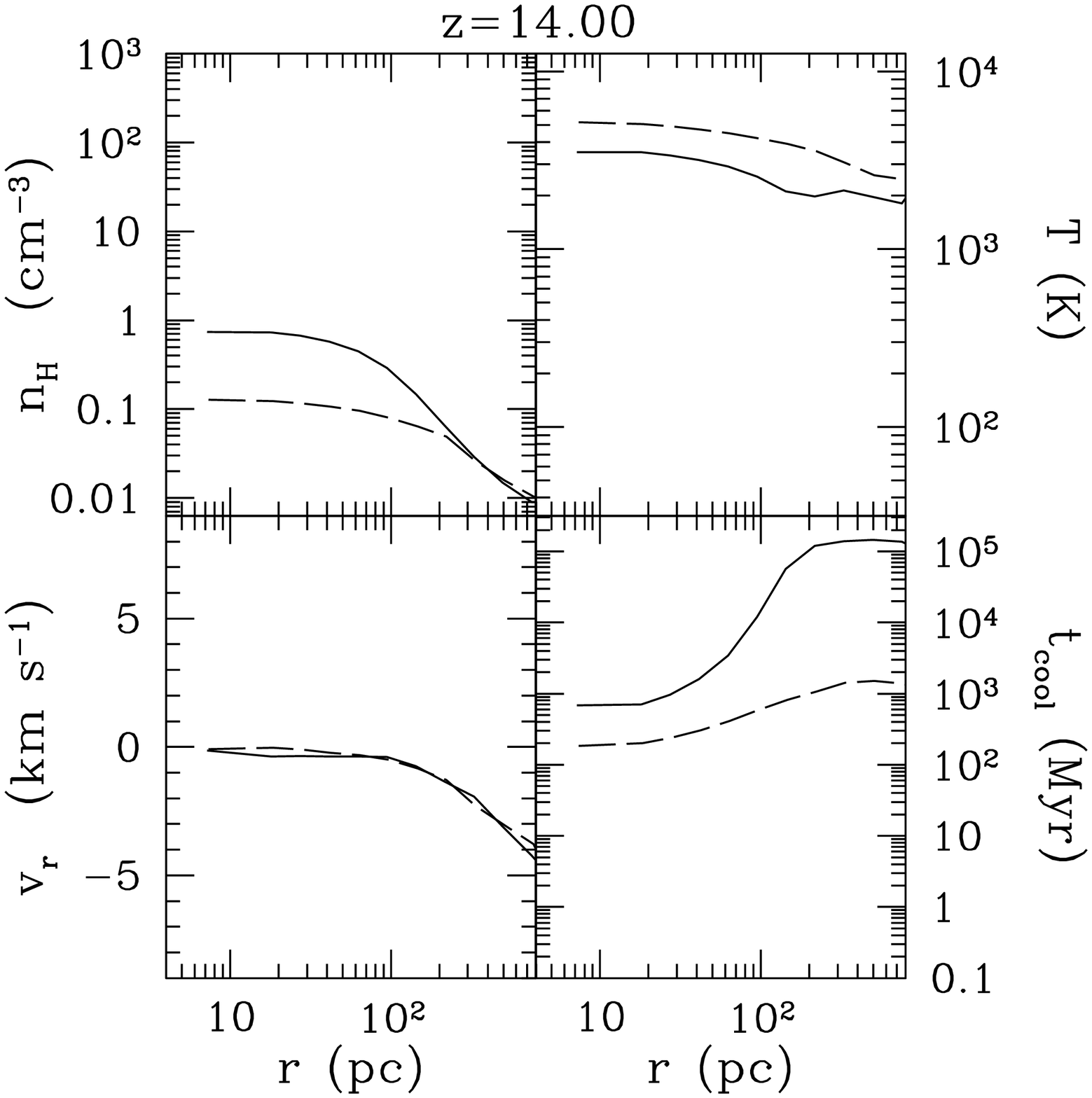}
\includegraphics[width=0.45\textwidth]{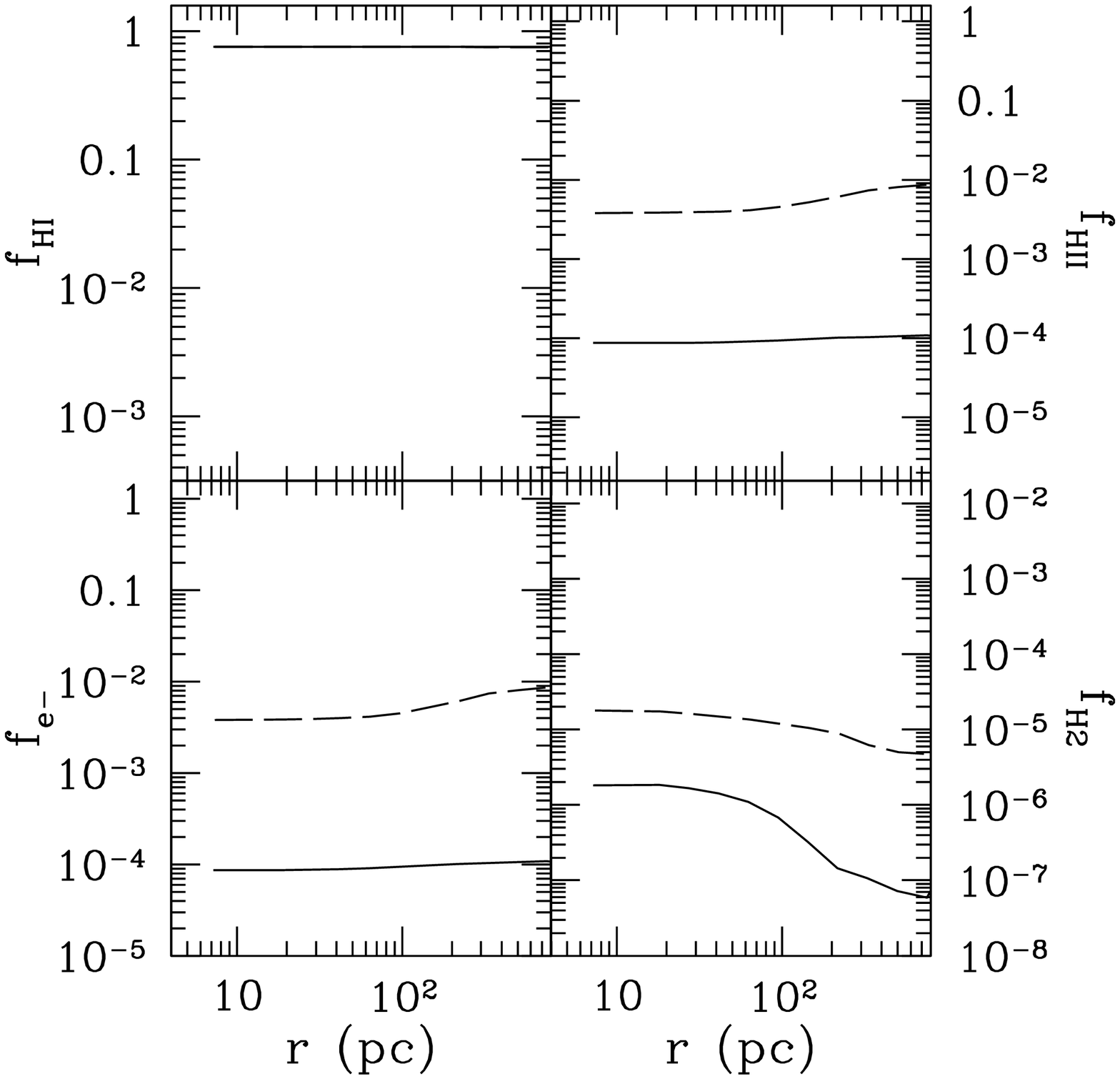}
 \caption{
Same as Fig. \ref{fig:NoUVB_vs_Heat}, at $z=14$, but with solid and dashed curves corresponding to the \LWonenothing\ and \LWoneheatinghigh\ runs, respectively.
\label{fig:LW1NoUVB_vs_LW1Heat}
}
\vspace{-1\baselineskip}
\end{figure}

Does this trend continue as we increase the strength of the UVB?  In Fig. \ref{fig:LW2NoUVB_vs_LW2Heat}, the solid and dashed curves correspond to the \LWtwonothing\ and \LWtwoheatinghigh\ runs at $z=14$, respectively; similarly in Fig. \ref{fig:LW1NoUVB_vs_LW1Heat}, the solid and dashed curves correspond to the \LWonenothing\ and \LWoneheatinghigh\ runs at $z=14$, respectively.  Here we see that the core density in the \LWoneheatinghigh\ run is lower than in the \LWonenothing\ run, but the cooling time is {\it also lower} by the same factor of $\sim2$ seen in Fig. \ref{fig:LW3NoUVB_vs_LW3Heat}, sourced by the enhanced H$_2$ fraction.  In fact, from higher redshift output and the previous figures, we see that the density is still ``catching up'' as time progresses; namely, the dashed density curve keeps getting closer to the solid density curve.  Thus we see that {\it halos inside relic HII regions, when exposed to the same LWB, always have a higher H$_2$ abundance and shorter cooling times than halos outside relic HII regions, allowing gas to cool faster once it finally begins to collapse onto the halo}.

It is important to note that in this regime where a strong LWB dominates the overall feedback, star formation is likely to be substantially delayed.  This delay can be long enough to allow the halo to be photoevaporated during reionization, before it had a chance to host additional stars.  Thus if one is only concerned with the fate of those minihalos contributing their ionizing photons to the progress of reionization, it is possible that the sole {\it persistent} feedback mechanism is {\it positive}.

\section{The Impact of Neglecting Radiative Transfer}
\label{sec:ss}

As we have noted a number of times in this paper, our simulations assume the optically thin limit, for both ionizing radiation and photodissociating LW photons.  Both of these approximations clearly fail at points in our simulations, and so it is necessary to understand the nature and implications of this approximation.  For clarity, we discuss ionizing photons and LW photons separately.

We discuss the impact of radiation hydrodynamic (RHD) effects due to ionizing photons first.  There is now a substantial body of literature on the ionization of primordial gas halos \citep{SIR04, WAN04, ISR05, ABS05, SU06, Whalen08, AWB07, WA08} and the broad outline of how halos are ionized is now well understood.  We begin by noting that the optical depth of the halos which are ionized in our simulation box at $z=25$ (as shown in Fig.~\ref{fig:early_profiles}), as measured from the virial radius to the center is -- at the Lyman limit -- well over 100.  Therefore, at first blush, the optically thin limit seems like a poor approximation.  However, this impression is not entirely accurate for two reasons.  

First, we are studying the {\em later} formation of halos in relic HII regions, and much of the gas that eventually falls into a halo at late times is {\em not} in the center of halos at reionization.  This is the question that \citet{OH03} asked: does the excess entropy generated in low-density gas in relic HII regions prevent it from forming stars at later times.  This low density gas has much lower optical depth and so is very likely to be ionized by a nearby Pop III star.  

Second, even for the gas in the centers of our halos at $\zuvbon$, the optically thin approximation is not as bad as the large optical depth at the Lyman limit indicates.  We can see this more clearly by comparing our results to RHD calculations.  For example, take the (typical) halo shown in Fig.~\ref{fig:early_profiles}.  The central density when ionization begins (at $z=25$) is just slightly over 1 cm$^{-3}$.  This value is much below the central density ($10^4$ cm$^{-3}$) assumed by \citet{SU04}, and is not that different from their critical density for self shielding (depending on the assumed distance of the source).  A better comparison is with the ``023" series of simulations run in \citet{Whalen08}, which have almost identical central densities.  As is shown in Fig. 17 of that paper, this halo is disrupted and dissociated for all assumed distances of the first star.  Comparing Fig. 9 and 10 of that paper to our Fig.~\ref{fig:early_profiles} shows some obvious differences due to radiative transfer effects, but the basic evolution is not all that different.  In particular, after a few Myr, the density at the halo center has decreased by about a factor of $\sim 5$ after $\sim 3$ Myr, while the molecular hydrogen fraction has increased to a few times $10^{-3}$.  These numbers are very similar to what we find, and it is these basic facts which, we argue, cause the delay in future star formation, as discussed earlier.  The reason for the similarity is that, as it sweeps over the halo, the ionization front can quickly ionize hydrogen until it transitions to a D-type front and stalls.  However, this happens only $\sim 15$ pc from the core, at densities slightly below 1 cm$^{-3}$, and the shock generated by the D-type front quickly disperses the core.

Of course, we do not claim to reproduce the radiation hydrodynamics in detail.  For example, we do not reproduce the thin self-shielded filament at the lower right of Fig. 10 of \citet{Whalen08}, nor do we follow the detailed shock-compression or "rocket" offset of the core.   However, by choosing to apply the ionizing flux at a redshift when the halos have low masses and central densities, we minimize the impact of the radiation transfer.  To verify that this is actually true in our simulations, we examined the central densities and masses of our halos at $z=25$; we find that all halos except for one have masses below $10^6$ $\Msun$ and central densities below $\sim$ 2 cm$^{-3}$ (one halo has $M > 10^6 \Msun$ and $n_c = 7 \times 10^3$ cm$^{-3}$).  Our halo abundance is consistent with predictions from the conditional Press-Schechter formalism (MBH06).  We chose this approach because we are interested in investigating the impact of relic HII regions on the formation of structure which forms significantly later, rather than studying the impact of a Pop III star on a nearby, highly collapsed halo.  In particular, as noted in the introduction, our motivation is to follow up on the suggestion of \citet{OH03} that relic HII regions will imprint an excess entropy on the low-density gas at early times, which will then suppress later star formation.  Of course, we stress that it is still important to examine our results with a more realistic RHD treatment. 

The results discussed in the section on positive feedback (section~\ref{sec:pos}) are even more robust because in that case the halos form out of gas that was barely overdense when the ionization occurred.  As can be seen in Fig.~\ref{fig:NoUVB_vs_Heat}, even at $z=21$ (long after the transient ionization), the central density of the halo (in the noUVB case) is only 0.2 cm$^{-3}$.

Another useful comparison of our results can be made with \citet{WA08b}, who used a full cosmological simulation of Pop III star formation including radiative transfer (albeit in a region considerably smaller than considered there).   While a direct comparison is difficult because \citet{WA08b} do not explicitly measure the suppression due to photo-ionization, a number of conclusions can be drawn.  First, it is clear from Fig. 7 of that paper that they also see the evaporation of gas from low-mass halos and filaments, and the resulting density field in the two simulations looks remarkably similar.  In addition, the time difference between the first Pop III object and the formation of the second is about 80 Myr, consistent with the time-delay derived in this paper.  On the other hand, one important point that comes out of that paper is that, because Pop III star formation tends to be clustered, relic HII regions do not sit isolated for longer periods of time and so a region of space may undergo multiple ionization episodes, something which is not considered in this work.

Next, we turn to LW photons.  First, we remind the reader that we are discussing a persistent LW background, and that although the LW flux is produced by the same {\it class} of objects that generate the transient UV flux, it is not produced by the same physical objects.  As discussed in more detail in \citet{HRL96}, the LW background is accumulated from a large number of distant sources. 
The effective optical depth in the LW lines in the background IGM,
over a Hubble distance, is at most $\tau\approx 1-2$
\citep{HAR00,RGS01}, so the Olbers' integral for the LWB is dominated
by the large number of sources at a good fraction of the Hubble
distance.  In comparison, the ionizing photons come exclusively from
the much fewer source(s) nearby, within the local HII region.

LW photons can be self-shielded by a sufficient column of H$_2$.  Here again, there are two regimes.  One is the large quantities of low-density gas in the relic HII regions that form substantial amounts of molecular hydrogen immediately after the ionizing flux is turned off (e.g. \citealt{KM05}).  Although H$_2$ formed in this manner may slow the build-up of a persistent LW background, it does not -- for a given LW background -- provide a large amount of shielding.  For example, \citet{JGB07} find typical values of LW optical depths of order unity for a single relic HII region.   This means that the LW flux inside the relic HII region can be reduced by a factor of a few, but it also means that it doesn't take long for the persistent LW background to "burn off" the H$_2$ produced in relic HII regions. 

The second regime occurs after the ionized gas falls back into the collapsed halos, where H$_2$ is in equilibrium between formation and photo-destruction.  In this case, when a sufficient column density of H$_2$ is formed, it can effectively attenuate the LW flux.  Based on our simulations, we find that when CD gas forms, the LW optical depth can be of order unity or larger, indicating that self-shielding can play a role during the recollapse (although probably not during the photoionization process itself).  Therefore, we caution readers that the critical LW values we give above may need to be increased when self-shielding is properly accounted for.

Finally, we note that in simulations with full radiative transfer, the photoionizing flux can also vary across a given halo (relic H II regions have radii of 2-3 kpc, while the Lagrangian radius of a typical halo in our simulations is of order 0.5 kpc).  This also means that, while most forming halos lie either completely inside or outside a relic H II region, some proto-halos could be partially ionized.


\section{Conclusions}
\label{sec:conc}

UV radiation from early astrophysical sources could have a large impact on subsequent star formation in nearby protogalaxies, and in general on the progress of cosmological reionization.  Theoretical arguments based on the absence of metals in the early Universe suggest that the first stars were likely massive, bright, yet short-lived, with lifetimes of a few million years.  Here we study the impact of the transient radiation arising from such stars on early protogalaxies.  We apply the same statistical approach as in MBH06, studying various combinations of transient UVBs and persistent LWBs, using the hydrodynamical simulation code, Enzo.  We also study a more typical and relevant region whose proto-galaxies form at lower redshifts, $z\sim$ 13--20. This allows us to trace feedback effects on longer time-scales and lower redshifts.

We confirm the results of MBH06 that feedback in the relic HII regions resulting from such transient radiation, is itself transient.  Feedback effects dwindle away after $\sim30\%$ of the Hubble time, and the same critical specific intensity of $J_{\rm UV} \sim 0.1 \times 10^{-21}{\rm ergs~s^{-1}~cm^{-2}~Hz^{-1}~sr^{-1}}$ separates positive and negative feedback regimes. A weaker UVB stimulates subsequent star formation
inside the relic HII regions, by enhancing the molecular hydrogen abundance. A stronger UVB delays star-formation by reducing the gas density at the centers of collapsing halos. The fact that we have confirmed the results in MBH06 (which were mostly inferences obtained by extrapolating from higher redshifts since we were unable to continue those simulations to low enough redshifts) suggests that overall feedback is fairly  insensitive to the large-scale environment, overdensity, and redshift-dependent halo parameters, and can accurately be modeled in this regime with just the intensity of the impinging UVB.

 We note that a value of $\Jlwb\sim 10^{-3}$--$10^{-2}$ separates feedback regimes dominated by the LW background from those dominated by the transient UVB.  As discussed in section~\ref{sec:ss}, there is some uncertainty in these values due to our optically thin treatment.  In particular, accounting for LW self-shielding may increase the values of $\Jlwb$ quoted here.

Additionally, we discover a second episode of eventual positive feedback in halos which have not yet collapsed when their progenitor regions were exposed to the transient UVB.  
Gas in such late-forming objects did not have time to set-up a steep density profile, which would translate into a strong pressure shock when photo-heated.  Instead they formed out of material which was of more moderate densities at $\zuvbon$ where the lasting impact of the transient UVB was positive: seeding the gas with an enhanced abundance of molecular hydrogen, which aids in its eventual cooling.
However, this feedback regime is very sensitive to the presence of Lyman-Werner radiation, and disappears under fairly modest background intensities of $\JLW \gsim 10^{-3} \times 10^{-21}{\rm ergs~s^{-1}~cm^{-2}~Hz^{-1}~sr^{-1}}$.
Nevertheless, when exposed to the same LWB, halos inside relic HII regions always have a higher H$_2$ abundance and shorter cooling times than halos outside relic HII regions, allowing gas to cool faster once it finally begins to collapse onto the halo.
  Although it is difficult to accurately estimate the build-up of the LWB, it seems likely that such modest values were surpassed well before the bulk of reionization \citep{HAR00}.  Thus this persistent eventual positive feedback might only be of academic interest, provided LW self-shielding is not efficient.

We conclude that radiative feedback from a short-lived UVB seems unlikely to have a major impact on the progress of global cosmological reionization, provided that present estimates of the luminosities and lifetimes of Pop III stars are accurate (e.g. \citealt{Schaerer02}).  Thus it is likely that the LWB plays the dominant role in regulating feedback in relic HII regions.

\vskip+0.5in

We thank the referee for comments which significantly improved the presentation of this paper.  Support for this work was provided by NASA through Hubble Fellowship grant \#HF-01222.01 to AM, awarded by the Space Telescope Science Institute, which is operated by the Association of Universities for Research in Astronomy, Inc., for NASA, under contract NAS 5-26555.  GB acknowledges support from NSF grants AST-05-07161, AST-05-47823, and AST-06-06959, as well as computational resources from the National Center for Supercomputing Applications.
ZH thanks the Pol\'{a}nyi Program of the Hungarian National Office of Technology.

\bibliographystyle{mn2e}
\bibliography{ms}

\end{document}